\documentclass[11pt]{article} 

\usepackage{amsmath,amsthm,latexsym,amssymb,amsfonts,epsfig}

\newcommand{\be}{\begin{equation}}
\newcommand{\ee}{\end{equation}}
\newcommand{\ba}{\begin{eqnarray}}
\newcommand{\ea}{\end{eqnarray}}

\title{{\sf Covariant Origin of the}\\
 {\sf $U(1)^3$ model for Euclidean Quantum Gravity}} 
\author{
{\sf S. Bakhoda}$^{1,2}$\thanks{{\sf 
s\_bakhoda@sbu.ac.ir, sepideh.bakhoda@gravity.fau.de}},
{\sf T. Thiemann}$^2$\thanks{{\sf 
thomas.thiemann@gravity.fau.de}}\\
\\
{\sf $^1$ Department of Physics, Shahid Beheshti University, Tehran, Iran}\\
{\sf $^2$ Institute for Quantum Gravity, FAU Erlangen -- N\"urnberg,}\\
{\sf Staudtstr. 7, 91058 Erlangen, Germany}\\
}
\date{{\small\sf \today}}

\makeatletter
\@addtoreset{equation}{section}
\makeatother

\begin{document} 

\maketitle

{\sf

\begin{abstract}
If one replaces the constraints of the Ashtekar-Barbero $SU(2)$ gauge theory 
formulation of Euclidean gravity by their $U(1)^3$ version, one arrives at a consistent model which captures significant structures of its $SU(2)$ version.
In particular, it displays a non-trivial realisation of the hypersurface 
deformation algebra which makes it an interesting testing ground for (Euclidean)
quantum gravity as has been emphasised in a recent series of papers due 
to Varadarajan et al. 

The simplification from SU(2) to U(1)$^3$ can be performed simply by hand 
within the Hamiltonian formulation by dropping all non-Abelian 
terms from the Gauss, spatial diffeomorphism, and Hamiltonian 
constraints respectively. However, one may ask from which Lagrangian 
formulation this theory descends. For the SU(2) theory it is known 
that one can choose the Palatini action, Holst action, or (anti-)selfdual action (Euclidean signature) as starting point all leading to equivalent
Hamiltonian formulations.

In this paper, we systematically analyse this question directly for the 
U(1)$^3$ theory. Surprisingly, it turns out that the Abelian 
analog of the Palatini or Holst formulation is a consistent but topological 
theory without 
propagating degrees of freedom. On the other hand, a twisted Abelian analog of the (anti-)selfdual formulation does lead to the desired Hamiltonian
formulation.

A new aspect of our derivation is that we work with 1. half-density valued tetrads which simplifies the analysis, 
2. without the simplicity constraint (which admits one undesired solution that is usually neglected by hand) and 3. without imposing the time gauge from the beginning.
As a byproduct, we show that also the non-Abelian theory admits a twisted 
(anti-)selfdual formulation. Finally, we also derive a pure connection 
formulation of Euclidean GR including a cosmological constant 
by extending previous work due to 
Capovilla, Dell, Jacobson, and Peldan which may be 
an interesting starting point for path integral investigations and 
displays (Euclidean) GR as a Yang-Mills theory with non-polynomial 
Lagrangian. 
\end{abstract}

\section{Introduction}
\label{s1}

In our two companion papers \cite{0} we studied 
the reduced phase space formulation of the SU(2) $\to$ U(1)$^3$  
truncation model for Euclidean signature General Relativity introduced 
in \cite{0a}. The motivation is to set up a reduced phase space 
quantisation as an alternative to the operator constraint quantisation 
much studied recently in \cite{7}. The two quantisation methods have 
complementary advantages and disadvantages: The reduced approach allows
to work directly with gauge-invariant variables, the physical Hilbert space 
and a physical time evolution while in the constrained approach these 
can only be extracted after having found the full solution space of the 
quantum constraints. On the other hand, the reduced approach in general 
uses suitable gauge fixings such that manifest covariance is lost and 
it requires non-trivial effort to prove that different gauges 
result in quantum theories that make identical predictions for gauge 
invariant objects, also gauge fixings may not be perfect and suffer from 
the existence of Gribov copies. In general, one expects that the 
two approaches enrich each other by providing complementary insight.
In particular, the programme of \cite{7} is designed to reduce 
quantisation ambiguities of the physical Lorentzian theory \cite{1,2,3,5}
in the framework of Loop Quantum Gravity (LQG), \cite{4} using the simpler 
U(1)$^3$ model as a test laboratory and we aim to support that enterprise 
with our reduced phase space approach.

In our companion papers, \cite{0} 
we focussed on the Hamiltonian formulation of the theory which is well suited as a starting point for a canonical quantisation of the reduced theory. In this paper, we consider the Lagrangian formulation of U(1)$^3$ 
theory where we have a subsequent path integral quantisation in mind. 
Note that the U(1)$^3$ theory is usually introduced as a truncation by hand of the Hamiltonian formulation of Euclidean General Relativity and not derived from an action principle. Thus the question arises whether such 
actions exist and how they look like. 

The answer is not a priori clear: How should the gauge group
of the Lagrangian be chosen? For Euclidean GR one can start with the 
SO(4) Palatini-Holst action \cite{9}
and after a tedious constraint analysis involving second 
class constraints\footnote{It is worth mentioning that in the first reference of \cite{15} the second-class
constraints of the Holst action are solved in a manifestly Lorentz-covariant way and in the second one the authors performed the canonical analysis of the Holst action using an
appropriate parametrisation of the tetrad and the connection. Finally, they arrived at
the Hamiltonian formulation of the Holst action which involves only first-class constraints after
integrating out some auxiliary fields. Their procedure avoids the
introduction of second-class constraints.} 
and using the time gauge to fix the ``boost'' part of 
SO(4) (which is one of the two copies of SU(2) into which it factorises
modulo sign issues) one arrives at the usual Hamiltonian SU(2) formulation.
Thus, a natural guess could be that a suitable covariant action starts 
from a suitably ``abelianised'' version of the SO(4) Palatini Holst action, 
perhaps SU(2)$\times$U(1)$^3$ or U(1)$^6$. 
In \cite{0a}, U(1)$^3$ itself was used to propose an action that should
serve as a covariant origin of the Hamiltonian U(1)$^3$ model. We could 
not find a proof in the literature that this action indeed serves this purpose.
In appendix \ref{appendix2} we  show  that the action of \cite{0a} has too many degrees of 
freedom (it has 6 rather than 4 propagating degrees of freedom). This is 
maybe not too surprising because up to the replacement of SU(2) by U(1)$^3$ the action of \cite{0a} is 
the same as that of \cite{10} where an explicit analysis also revealed 6 propagating 
degrees of freedom. We therefore have  to  look  for  a  different covariant action.
The motivation to look at U(1)$^6$ was that it is the natural compact Abelian group of the same dimension as compact SO(4), analogous as U(1)$^3$ is the natural Abelian group of the same dimension as compact SO(3) or SU(2). Then boiling it down from U(1)$^6$ to U(1)$^3$ is supposed to happen in the time gauge.   

These kinds of actions would be of ``Palatini-Holst'' type. Yet another very interesting possibility is that one can find a pure connection formulation 
of the theory without using tetrads at all. That such a possibility in 
principle exists in presence of a 
cosmological constant
was pointed out in the context of the original self-dual 
theory for Lorentzian signature in works by Capovilla, Dell and 
Jacobson \cite{8} and Peldan \cite{11}, however, no closed formula 
was provided. 
In fact, some recent and in-depth analysis of pure connection formulations after Peldan was promoted in \cite{14, 16}. Nevertheless, here our aim is not to derive the most general action for the U(1)$^3$ theory but rather give a concrete example where such a non-polynomial pure connection Lagrangian can be given explicitly.
Such a Lagrangian would necessarily be rather non-polynomial 
in the curvature of the connection but spacetime diffeomorphism covariant
and is interesting in its own right because it would put (Euclidean) 
General Relativity on equal footing with (Euclidean) Yang-Mills theory
in the sense that only a connection is required to formulate the action.\\
\\
The architecture of this article is as follows:\\
\\
In section two we introduce the notation and collect the formulae for the Hamiltonian formulation of SU(2) or U(1)$^3$ Euclidean General Relativity.
The presentation will be brief, for more details we refer to \cite{0} and references therein. 

In section three we first consider the U(1)$^6$ Palatini action as a possible covariant origin of the U(1)$^3$ model. Surprisingly, this theory 
is {\it topological}, it has no propagating degrees of freedom. 
The technical reason for this striking difference with the SO(4) theory 
is that in the Dirac constraint stability analysis in a crucial step 
the Abelian nature of the model forbids solving the equations for 
the existing Lagrange multipliers but rather results in additional secondary
constraints. We then consider a one-parameter family of Lagrangians 
based on the gauge group U(1)$^3$. The parameter plays a role similar to 
but different from the Immirzi parameter and gives rise to a ``twisted
self-dual'' and Abelian connection. 
We show that in this case the Hamiltonian theory
is indeed equivalent to the U(1)$^3$ truncation of Euclidean General 
Relativity. As a byproduct, we show that the same action for 
SU(2) rather than U(1)$^3$ also gives rise to the Hamiltonian formulation 
of Euclidean General Relativity which so far was only known for the 
Euclidean (anti-)self-dual theory (the parameter equals plus or minus unity).
Note that despite starting from an SU(2) theory, this theory is different from 
the Husain-Kuchar theory \cite{10} as our model has a non-trivial 
Hamiltonian constraint. As a technical advance, we perform the analysis 
using half densitised tetrads and avoiding the introduction of simplicity 
constraints which always have spurious solutions.

In section four we review the programme of \cite{8,11} and carry out the required computations to the end. We show that the final equation to be solved for the backward Legendre transform can be brought into a rather manageable and explicit 
form and can be stated as the roots of a quartic polynomial whose 
coefficients are simple polynomials in the spacetime covariant curvature.
The roots are algebraically accessible using the Cardano-Ferrari theory 
\cite{12} and will be analysed further in future publications.

In section five we conclude and point out directions for further work. 

In appendix \ref{appendix}, we analyse the general solution of an equation that 
arises in the Dirac stability analysis of section three. 

In appendix \ref{appendix2}, we present the Hamiltonian analysis of the Somlin's action \cite{0a} and conclude that it does not lead to the Hamiltonian U(1)$^3$ model. 
\section{Notation and review of the Hamiltonian 
U(1)$^3$ model for Euclidean vacuum GR}
\label{s2}               

We use $\mu,\nu,\rho,...\in \{0,1,2,3\}$ 
as general spacetime tensor indices and 
$A,B,C,..\in \{t,1,2,3\}$ as those adapted to an ADM slicing. Likewise we will
use $a,b,c,...\in \{1,2,3\}$ as spatial ADM indices. Furthermore 
$I,J,K,...\in \{0,1,2,3\}$ is a SO(4) index and 
$i,j,k,...\in \{1,2,3\}$ is an SU(2) or $U(1)^3$ index.

By $N,N^a, \Lambda^j$ we denote respectively lapse function, shift vector and 
Lagrange multiplier of the SU(2) or U(1)$^3$ Gauss constraint. The canonically
conjugate variables are $A_a^j$ and $E^a_j$. The difference between the 
U(1)$^3$ truncation and the SU(2) model arise in the Gauss constraint and the 
curvature which for SU(2) read 
\be \label{2.1} 
G_j=\partial_a E^a_j+\epsilon_{jkl} A_a^k E^a_l,\qquad
F_{ab}^j=2\partial_{[a} A_{b]}^j+ \epsilon_{jkl} A_a^k A_b^l
\ee
while for U(1)$^3$ the terms proportional to $\epsilon_{jkl}$ are missing.

The spatial diffeomorphism and Hamiltonian constraint (with cosmological 
constant) are identical for both models and read
\be \label{2.2}
C_a=F_{ab}^j\; E^b_j,\qquad
C=[\det(E)]^{-1/2}[F_{ab}^j\epsilon_{jkl} E^a_k E^b_l+\Lambda \det(E)]
\ee

\section{Covariant U(1)$^6$ theory}
\label{s3}

In this section we consider an Abelian analog of the Palatini or Holst 
action as a potential Lagrangian formulation of the U(1)$^3$ theory. 
In the first subsection we recall the procedure that one follows in the 
non-Abelian case. We use it as a guideline to carry out the corresponding 
analysis in the Abelian case in the second subsection.  

\subsection{Review: SO(4) Holst action}
\label{s3.1}

The non-Abelian SU(2) formulation can be obtained from the Palatini action 
or the Holst action for the gauge group SO(4). The two actions differ by 
a topological term multiplied by the Immirzi parameter \cite{9}. The way 
this works is that the momentum conjugate to the SO(4) connection 
has to descend from a tetrad which we call {\it tetrad constraint} 
$T^{ab},\; a,b=1,2,3,\; T^{[ab]}\equiv 0$. 
From the tetrad constraint one deduces the {\it simplicity constraint} 
$S^{ab}$
which has the advantage that one can formulate it purely in terms of 
of the momentum conjugate to the connection. Unfortunately the tetrad 
constraint and simplicity constraint are not equivalent: The simplicity 
constraint admits another solution inequivalent to the tetrad constraint
which is usually neglected by hand. In order to avoid any doubtful 
reasoning, we therefore avoid the simplicity constraint 
and work instead with the proper tetrad constraint.

Nevertheless, requiring the dynamical stability of the primary 
simplicity constraint
yields among others a secondary constraint 
which we call {\it dynamical constraint} $D^{ab}$.  
In the non-Abelian theory the Dirac algorithm stops here (there are no 
tertiary constraints), the 
constraints $S^{ab}, D^{ab}$ form a second class pair and the Lagrange 
multiplier of $S^{ab}$ within the primary Hamiltonian is therefore completely 
fixed by the stability condition. To arrive at the Hamiltonian formulation 
in terms of SU(2) one utilises the Lie algebra isomorphism 
so(4)$\cong$su(2)$\oplus$su(2) and gauge fixes one copy of the decoupled 
su(2) Gauss constraints using the time gauge $e^t_j=0$ ($e^A_I$ is the 
tetrad with tensorial indices $A=t,a;\; a=1,2,3,$ and Lie algebra indices
$I=0,j;\; j=1,2,3$). The Hamiltonian formulation then results from solving 
the second class constraints, half of the Gauss constraints and the time 
gauge and by passing to the corresponding Dirac bracket.

\subsection{The U(1)$^6$ model}
\label{s3.2}

As we have just seen, in the non-Abelian case the Hamiltonian 
su(2) formulation results from the Lagrangean so(4) formulation by 
gauge fixing one of the su(2) copies. In the Abelian case this suggests 
to start from the Lie algebra 
$\mathfrak{g}=$u(1)$\oplus$u(1)$\oplus$u(1)$\oplus\mathfrak{h}$
where $\mathfrak{h}$ is some Lie algebra such that $\mathfrak{g}$
admits a four dimensional representation (in which the tetrad transforms 
while the connection transforms in the adjoint). If we wish to stay as close 
as possible to the non-Abelian theory then $\mathfrak{g}$ should be 
six dimensional. If the gauge group corresponding to $\mathfrak{g}$ is 
supposed to be compact then the one corresponding to $\mathfrak{h}$ is also 
compact. Finally, if $\mathfrak{g}$ is supposed to be Abelian this 
fixes $\mathfrak{g}$ as direct sum of six u(1) copies. 

Accordingly, the Lagrangian variables are a U(1)$^6$ connection 
$A_B^{IJ},\; A_B^{(IJ)}=0$ and a tetrad $e^A_I$ which transforms 
in the trivial representation of U(1)$^6$.  
The expectation is then that in analogy to the non Abelian case 
three copies of U(1) get gauge fixed in the course of the Dirac algorithm,
either by hand as in the non-Abelian case or as a consequence of the 
stability conditions. In the first case it is clear that the time
gauge cannot serve as gauge fixing condition since the tetrad is 
Gauss invariant. In the second case the time gauge could arise as a second 
class constraint.

Surprisingly, while consistent, the U(1)$^6$ model turns out to be 
topological. As in the non-Abelian case, the Dirac analysis stops at the 
secondary level, i.e. there are no tertiary constraints. The secondary 
constraints involve the Abelian analog of the dynamical constraint $D^{ab}$.
However, precisely due to the Abelian nature of the model, the pair 
of constraints $(T^{ab}, D^{ab})$ is now first class rather than second 
class. By the ususal naive counting, this leads to a reduction by 24 rather 
than 12 degrees of freedom. As the non-Abelian theory only has 4 propagating
degrees of freedom, this means that the resulting theory is topological.\\
\\
In what follows we provide the details of the analysis. We start from the
action 
\be \label{3.1}
S=\frac{1}{2}\;\int\;dt\;d^3x\; F_{AB}^{IJ}\; \hat{\sigma}^{AB}_{IJ},\qquad
\hat{\sigma}^{AB}_{IJ}=\hat{\Sigma}^{AB}_{IJ}+\frac{1}{2}
\gamma \epsilon_{IJ}\;^{KL}
\hat{\Sigma}^{AB}_{KL},\qquad \hat{\Sigma}^{AB}_{IJ}=\hat{e}^A_{[I} \;\hat{e}^B_{J]}
\ee
where Lie algebra indices are moved with the Kronecker symbol. Here we work
with half density valued tetrads 
\be \label{3.2}
\hat{e}^A_I=\det(\{e_B^J\})^{1/2}\; e^A_I
\ee
in terms of which the action is polynomial. It is assumed that the tetrad is 
nowhere degenerate so that w.l.g. its determinant is everywhere positive.
Here $\gamma$ is the Immirzi parameter. In this subsection we assume 
$\gamma\not=\pm 1$ because otherwise the connection is projected 
onto its (anti-)selfdual part. The values $\gamma=\pm 1$ 
will be treated in the next subsection as a special case. The analysis for 
$\gamma\not=\pm 1$ is qualitatively similar to the case $\gamma=0$ so that 
we set $\gamma=0$ for the rest of this subsection.

The 3+1 decomposition reveals 
\be \label{3.3}
S=\int\;dt\; d^3x\; [F_{ta}^{IJ}\; \hat{\Sigma}^{ta}_{IJ}+
\frac{1}{2}\;F_{ab}^{IJ}\; \hat{\Sigma}^{ab}_{IJ}]
\ee
Computing the momenta $\pi^B_{IJ},\; \hat{P}_A^I$ conjugate to 
$A_B^{IJ},\;\hat{e}^A_I$ we find the primary constraints
\be \label{3.4}
\pi^t_{IJ}=0,\qquad T^a_{IJ}:=\pi^a_{IJ}-\hat{\Sigma}^{ta}_{IJ}=0,\qquad \hat{P}_A^I=0
\ee
and the Legendre transform of (\ref{3.3}) yields with the velocities 
$v_a^{IJ},\hat{v}^A_I$ that one cannot solve for 
the primary Hamiltonian
\ba \label{3.5}
H &=&\int\; d^3x\; \{v_A^{IJ}\; \pi^A_{IJ}+\hat{v}^A_I \hat{P}_A^I-L\}
\nonumber\\
&=& \int\; d^3x\; \{v_a^{IJ}\; T^a_{IJ}+\hat{v}^A_I \hat{P}_A^I
+v_t^{IJ}\;\pi^t_{IJ}+A_{t,a}^{IJ}\; \hat{\Sigma}^{ta}_{IJ}-\frac{1}{2}
F_{ab}^{IJ}\;\hat{\Sigma}^{ab}_{IJ}\}
\ea
Stability of $\pi^t_{IJ}=0$ yields the U(1)$^6$ Gauss secondary constraint
\be \label{3.6}
G_{IJ}=\partial_a\; \hat{\Sigma}^{ta}_{IJ}
\ee  

Stability of $\hat{P}_A^I=0$ yields with the Lagrange multiplier 
$f_a^{IJ}:=v_a^{IJ}-\partial_a A_t^{IJ}$ the 16 equations
\be \label{3.7}
f_a^{IJ}\; \hat{e}^a_J=F_{ab}^{IJ}\;\hat{e}^b_J-f_a^{IJ}\; \hat{e}^t_J=0
\ee
We solve them by decomposing $f_a^{IJ}=f_a^{BC}\;\hat{\Sigma}^{IJ}_{BC}$ where
$\hat{\Sigma}^{IJ}_{AB}=\hat{e}^{[I}_A\;\hat{e}^{J]}_B,\;
\hat{e}^A_I\;\hat{e}^I_B=\delta^A_B,\; 
\hat{e}^A_I\;\hat{e}^J_A=\delta^J_I$ 
and exploiting the relations
\be \label{3.8}
\hat{\Sigma}^{AB}_{IJ}\; \hat{\Sigma}^{IJ}_{CD}=\delta^A_{[C}\;\delta^B_{D]},
\qquad
\hat{\Sigma}^{AB}_{IJ}\; \hat{\Sigma}^{KL}_{AB}=\delta^K_{[I}\;\delta^L_{J]}
\ee
We find the following restriction on the Lagrange multipliers
\be \label{3.9}
f_a^{bt}=F_{ac}^{IJ}\;\hat{\Sigma}^{bc}_{IJ},\qquad 
f_a^{bc}=\epsilon^{bcd} s_{ad},\qquad s_{[ad]}=0
\ee
and the 4 secondary constraints
\be \label{3.10}
C_a:=F_{ab}^{IJ} \;\hat{\Sigma}^{tb}_{IJ},\qquad
C:=F_{ab}^{IJ} \;\hat{\Sigma}^{ab}_{IJ}
\ee
That is, 12 of 18 Lagrange multipliers have been fixed while the 6 degrees 
of freedom encoded by the symmetric tensor $s_{ab}$ remain free. The 
16 relations (\ref{3.9}) and (\ref{3.10}) ensure stability of $\hat{P}_A^I$.

Stability of $T^a_{IJ}$ yields
\be \label{3.11}
\partial_b\hat{\Sigma}^{ab}_{IJ}
+\hat{v}^t_{[I} \hat{e}^a_{J]}  
+\hat{e}^t_{[I} \hat{v}^a_{J]}=0
\ee  
We decompose $\hat{v}^A_I=\hat{v}^A_B\;\hat{e}^B_I$ and find 
the restriction on the Lagrange multipliers
\be \label{3.12}
\hat{v}^a_b=-[2\hat{\Sigma}^{IJ}_{tb}\;\hat{\Sigma}^{ca}_{IJ,c}+\hat{v}^t_t
\delta^a_b],\qquad
\hat{v}^t_a=-\hat{\Sigma}^{IJ}_{ab}\;\hat{\Sigma}^{cb}_{IJ,c} 
\ee
and the 6 secondary constraints
\be \label{3.13}
D^{ab}=\epsilon^{cd(a}\;\hat{\Sigma}^{b)e}_{IJ,e}\; \hat{\Sigma}^{IJ}_{cd}
\ee
That is, 12 of 16 Lagrange multipliers have been fixed while the 4 
multipliers $\hat{v}^A_t$ remain free. Altogether the 18 relations 
(\ref{3.12}), (\ref{3.13}) ensure stability of $T^a_{IJ}=0$. 

We now must stabilise $G_{IJ}, C_a, C, D^{ab}$. This is most conveniently
done as follows. We have modulo $T^a_{IJ}=0$ that 
$G_{IJ}=\partial_a\pi^a_{IJ}$. Since $T^a_{IJ}$ is already stabilised, we 
may equivalently stabilise the constraint in the form 
$\hat{G}_{IJ}:=\pi^a_{IJ,a}=0$ in which it generates U(1)$^6$ 
gauge transformations. 
Since $H$ depends on $A_a^{IJ}$
only through its curvature $F_{ab}^{IJ}=2\partial_{[a} A_{b]}^{IJ}$ the 
Gauss constraint $\hat{G}_{IJ}$ is already stabilised. Next, modulo
$T^a_{IJ}=G_{IJ}=0$ we have 
$C_a=F_{ab}^{IJ} \pi^b_{IJ}-A_a^{IJ} \hat{G}_{IJ}$ in which form it generates 
spatial diffeomorphisms on $(A_a^{IJ},\pi^a_{IJ})$. Since $\hat{P}_A^I$ 
has already been stabilised, we add terms linear in $\hat{P}_A^I$ to $C_a$
so that the resulting constraint $\hat{C}_a$ generates spatial 
diffeomorphisms also on the variables $\hat{e}^A_I,\hat{P}_A^I$ (in doing so 
explicitly, note the density weight $\pm 1/2$ respectively). Since the 
primary Hamiltonian only depends on constraints which are tensor densities,
the constraint $\hat{C}_a$ and thus $C_a$ is already stabilised.   
 
Unfortunately, these abstract arguments are not available for $C, D^{ab}$.
We show some stages of the surprisingly tedious computation
\ba \label{3.14}
\{H,C(f)\} &=& \int\; d^3x\;
\{v_a^{IJ} \pi^a_{IJ}+\hat{v}^A_I\;\hat{P}_A^I,C(f)\}
\nonumber\\
&=& 2\int\; d^3x\;f\;
[\hat{\Sigma}^{bc}_{IJ}\;\partial_{[b} v_{c]}^{IJ}+\hat{v}^b_K\;
\hat{e}^c_L\; F_{bc}^{KL}]
\nonumber\\
&=& 2\int\; d^3x\;f\;
[\hat{\Sigma}^{bc}_{IJ}\;\partial_{[b} f_{c]}^{IJ}+\hat{v}^b_K\;
\hat{e}^c_L\; F_{bc}^{KL}]
\nonumber\\
&=& 2\int\; d^3x\;f\;[(f_b^{ab})_{,a}-\hat{\Sigma}^{ab}_{IJ,a}\; f_b^{IJ}
+\hat{v}^b_t\; C_b+\hat{v}^b_a\; f_b^{at}]
\nonumber\\
&=& 2\int\; d^3x\;f\;[-\hat{\Sigma}^{ab}_{IJ,a}\; \hat{\Sigma}^{IJ}_{AB}\; 
f_b^{AB}+\hat{v}^b_a\; f_b^{at}]
\nonumber\\
&=& 2\int\; d^3x\;f\;[-s_{ab}\; D^{ab}\; + f_b^{tc}\;
(\hat{v}^b_c-\delta^b_c \;\hat{v}^t_t)+\hat{v}^b_a\; f_b^{at}]
\nonumber\\
&=& 2\int\; d^3x\;f\;\hat{v}^t_t\; C
\ea
where in the first step we isolated the contributing part of $H$, in the 
second step we carried out the Poisson brackets and integrated by parts,
in the third step we noticed that the r.h.s. depends on $v_a^{IJ}$ only
through $f_a^{IJ}$, in the fourth we inserted (\ref{3.9}), (\ref{3.12}),
in the fifth we dropped $C_b$ and used $f_b^{ab}=0$ and decomposed
$f_b^{IJ}$, in the sixth we used again (\ref{3.9}), (\ref{3.12}), and in the 
seventh we cancelled terms, dropped $D^{ab}$ and used $C=f_a^{at}$.   
Accordingly $C$ is stable where the precise form of $f_a^{IJ}, \hat{v}^A_I$ 
and the secondary constraints $C_a, C, D^{ab}$ had to be used in a crucial 
way.

In order to carry out the analysis for $D^{ab}$ it is useful to rewrite 
it in the simpler form
\be \label{3.15}
D^{ab}=\epsilon^{cd(a}\; w_c\;^{b)}\;_d,\qquad w_c\;^b\;_d=
\hat{e}^I_c \; \hat{e}^{b}_{I,d}
\ee
Using also the abbreviation $\sigma^a_b:=w_t\;^a\;_b=\hat{e}^I_t\;\hat{e}^a_{I,b}$ 
it is not difficult to check that 
\be \label{3.16}
\hat{v}^a_b=\sigma^a_b-[v^t_t+\sigma^c_c]\delta^a_b , \qquad
\hat{v}^t_a=-w_{(a}\;^b\;_{b)}
\ee
With this machinery we compute
\ba \label{3.17}
\{H,D^{ab}(g_{ab})\}
&=& \{\int\; d^3x\; \hat{v}^A_I\; \hat{P}_A^I,D^{ab}(g_{ab})\}
\nonumber\\
&=&
-\int\; d^3x\; g_{ab}\;\epsilon^{acd}\;[\hat{v}^t_c\; \sigma^b_d
+\hat{v}^e_c\; w_e\;^b\;_d+\hat{v}^b_t\; [\hat{e}^t_I \; \hat{e}^I_{c,d}]
-\hat{v}^b_{c,d}+\sigma^b_e\; w_c\;^e\;_d]
\nonumber\\
&=&
-\int\; d^3x\; g_{ab}\;\epsilon^{acd}\;[w_{(c}\;^e\;_{e)}\; \sigma^b_d
+\sigma^e_c\; w_e\;^b\;_d+\hat{v}^b_t\; [\hat{e}^t_I \; \hat{e}^I_{c,d}]
-\sigma^b_{c,d}+\sigma^b_e\; w_c\;^e\;_d]
\ea
where in the first step we isolated the contributing piece of $H$, in the 
second we carried out the Poisson bracket and integrated by parts and used 
above abbreviations, in the last step we realised that the piece 
$\propto \delta^a_b$ in $\hat{v}^a_b$ drops out due to symmetry of $g_{ab}$.
It is easy to verify
\be \label{3.18}
\epsilon^{acd}\; \sigma^b_{c,d}=-\epsilon^{acd}[\hat{e}^I_t\; \hat{e}^t_{I,d}
\sigma^b_c+\sigma^e_d\; w_e\;^b\;_c]
\ee
and with (\ref{3.15}) and decomposing into symmetric and antisymmetric piece 
wrt indices $a,e$
\be \label{3.19} 
\epsilon^{acd} w_c\;^e\;_d=D^{ae}+\epsilon^{aed} \; w_{[c}\;^c\;_{d]}
\ee
Dropping the term $\propto D^{ae}$ we can simplify
\be \label{3.20}
\{H,D^{ab}(g_{ab})\}
=\int\; d^3x\; g_{ab}\;[\sigma^b_c\; \epsilon^{acd} \; w_d-\hat{v}^b_t\;
u^a]
\ee
where 
\be \label{3.21}
w_a:=w_a\;^b\;_b+\hat{e}^I_t\; \hat{e}^t_{I,a},\qquad
u^a:=\epsilon^{acd} \hat{e}^t_I \; \hat{e}^I_{c,d}
\ee
The final step consists in verifying that 
\be \label{3.22}
u^a=-\frac{1}{2}\epsilon^{abc}\;\hat{\Sigma}^{IJ}_{bc}\; G_{IJ},\qquad
w_a=2\hat{\Sigma}^{IJ}_{ta}\; G_{IJ}
\ee
so that $D^{ab}$ is stable modulo $D^{ab},G_{IJ}$.\\
\\
Accordingly, there are no tertiary constraints. However, the reason for 
the absence of tertiary constraints is completely different as compared to 
the non-Abelian theory. There the absence of tertiary constraints 
was ensured by fixing Lagrange multipliers in the primary Hamiltonian while 
in the Abelian theory no Lagrange multipliers are fixed, the stability 
is ensured already by the secondary constraints.

This difference also leads to a crucial departure in the counting of the 
degrees of freedom\footnote{Note that a safe way to count degrees of freedom is to use Dirac's algorithm which 
in particular ensures that the constraints encountered are algebraically independent. 
Counting at the Lagrangian level can be tricky: for example the action \cite{10} 
reveals 24 configuration degrees of freedom, 4 diffeomorphism gauges and 3 
Yang-Mills type gauges. One cannot easily deduce from the Lagrangian that this 
theory has 3 propagating canonical pairs.}
between the Abelian and non-Abelian theories: In the 
U(1)$^6$ theory we have altogether $2 \cdot (4 \cdot 6 + 4 \cdot 4) = 80$ 
phase space 
degrees of freedom. We have $6+18+16=40$ primary constraints 
$\pi^t_{IJ}, T^a_{IJ}, \hat{P}_A^I$ and $6+4+6=16$ secondary constraints
$G_{IJ},C_a,C, D^{ab}$. Altogether 12 of the 18 $v_a^{IJ}$ and 12 of the 
16 $\hat{v}^A_I$ got fixed in the course of the stability analysis 
while 6 of the $v_a^{IJ}$ (encoded by $s_{ab}$) and 4 of the $\hat{v}_A^I$
(encoded by $\hat{v}^A_t$) remain free. Going through the same analysis 
as above one can verify that $\hat{C}$ is a first class constraint 
where $\hat{C}$ is the integrand of $H$. This implies that $C$ is also 
first class. Furthermore, the fact that $s_{ab},\; \hat{v}^A_t$ remain free 
implies that the $6+4$ constraints  $S^{ab}, \hat{P}_A$ multiplying them
are also first class. Explicitly
\be \label{3.23}
S^{ab}=\pi^{(a}_{IJ}\; \epsilon^{b)cd} \hat{\Sigma}^{IJ}_{cd},\qquad
\hat{P}_t=
\hat{P}_t^I \hat{e}^t_I-\hat{P}_a^I \hat{e}^a_I,\qquad
\hat{P}_a=
\hat{P}_a^I \hat{e}^t_I      
\ee
Finally $D^{ab}$ has weakly vanishing Poisson brackets with all 
constraints except $\hat{P}_A^I$. However, we may add a Term 
$\propto T^a_{IJ}$ to $D^{ab}$ such that the resulting $\hat{D}^{ab}$ 
has exactly vanishing Poisson brackets with all constraints. 
Specifically
\be \label{3.24}
\hat{D}^{ab}(g_{ab})
=D^{ab}(g_{ab})
+T^a_{IJ}(\hat{\Sigma}^{IJ}_{AB} h_a^{AB}(g)),\qquad
h_a^{ca}(g)=-g_{ab}\epsilon^{acd} \hat{e}^I_t \hat{e}^b_{I,d},\qquad
h_e^{tc}(g)=2\epsilon^{ad[c}\; w_e\;^{b]}_d\; g_{ab}
+\epsilon^{acd} g_{ae,d}
\ee
 
To summarise: We have 12 second class pairs formed by 12 out of 18 $T^a_{IJ}$ 
and 12 out of 16 $\hat{P}^A_I$ and 
$6+6+4+6+4+6=32$ first class constraints  
$\pi^t_{IJ},S^{ab},\hat{P}_A, G_{IJ},C, C_a, \hat{D}^{ab}$. Altogether these 
are $2 \cdot 12+32=56=40+16$ constraints which reduce $24+2 \cdot 32=88$ 
degrees of
freedom. In the non-Abelian theory, the analogs of $S^{ab}, D^{ab}$ 
form a second class pair. Thus there we have also 56 constraints but 
now we have 36 second class constraints and only 20 first class constraints
thus reducing only $2\cdot 20 +36=76$ degrees of freedom and leaving 4 propagating
ones. Accordingly, the U(1)$^6$ theory is consistent in the sense that 
the Dirac analysis does not lead to a contradiction, but it is topological 
in the sense that the constraint reduction leaves no local degrees of freedom.
The U(1)$^6$ theory therefore cannot be a Lagrangian origin of the 
Hamiltonian U(1)$^3$ theory which has 4 propagating degrees of freedom.

\subsection{The twisted selfdual model}
\label{s3.3}

The following analysis surprisingly applies to both U(1)$^6$ and SO(4) 
simultaneously. We consider the action 
\be \label{3.25}
S=\frac{1}{2}\;\int\;dt\;d^3x\; F_{AB}^{IJ}\; \hat{\Sigma}^{AB}_{IJ},\qquad
\hat{\Sigma}^{AB}_{IJ}=\hat{e}^A_{[I}\; \hat{e}^B_{J]}
\ee
but $F_{AB}^{IJ}$ is not an U(1)$^6$ or SO(4) curvature but rather a 
{\it twisted selfdual} U(1)$^6$ or SO(4) curvature, that is, 
\be \label{3.26}
F^{0j}=F^j=\frac{1}{2\gamma}\epsilon_{jkl}\; F^{kl}
\ee
with $F^j=2 dA^j$ and $F^j=2dA^j+\epsilon_{jkl} A^k\wedge A^l$ respectively 
a U(1)$^3$ and SU(2) curvature. Here $\gamma\not=0$ is similar
to but different from the 
Immirzi parameter: Note that condition (\ref{3.26}) defines the (anti-)selfdual
model only for $\gamma=\pm 1$. If one starts from the Holst action
(\ref{3.1}) 
with topological term and $\gamma=\pm 1$ one arrives at (\ref{3.25}) because 
then the standard curvature of SO(4) or U(1)$^6$  is projected into the curvature of the (anti-)selfdual (i.e. SU(2) or U(1)$^3$ respectively) connection. However, when 
$\gamma\not=\pm 1$ the connection stays a genuine SO(4) or U(1)$^6$ connection
and one does not arrive at (\ref{3.25}). Thus for $\gamma\not=\pm 1$ the action is new. Still, as we will 
demonstrate below, it turns out to be equivalent to Euclidean GR or its U(1)$^3$ truncation respectively.
The fact that the action contains 
a full 
co-tetrad $e_A^I$ rather than just three frame one forms $e_A^j$ 
makes it different from the SU(2) model defined by Husain and Kuchar 
\cite{10} which has no Hamiltonian constraint although we also have only 
a SU(2) (or U(1)$^3$) connection. It maybe puzzling how an SU(2) or U(1)$^3$ connection acts on $\mathbb{R}^4$ but 
this is by the same mechanism as a self-dual connection would act. Note that it is 
crucial that $\gamma\not=0,\infty$ as we would otherwise end up with the action of 
\cite{0a} which is known to have too many degrees of freedom (see appendix \ref{appendix2}).
\\
\\
The subsequent analysis turns out to be much simpler than in the previous 
subsection. To anticipate the essential result for readers not interested in 
the details, we sketch the outcome of the analysis: This time among the 
primary constraints we find 9 tetrad 
constraints $T^a_j=\pi^a_j-\sigma^{ta}_j$, 3 momentum constraints  
$\pi^t_j$ and the 16 momentum constraints $\hat{P}_A^I$ as before. 
The stability of $\pi^t_j$ enforces as secondary constraint the Gauss 
constraint $G_j$ for the corresponding gauge group. The stability of 
$\hat{P}_A^I$ leads to 7 secondary constraints $C_a, C, D_j$ 
where $C_a$ is the spatial diffeomorphism constraint, 
$C$ the Hamiltonian constraint
and $D_j=\hat{e}^t_j$ is the {\it time gauge constraint}. That is, the time 
gauge is {\it dynamically enforced} in this model and not a convenient 
gauge choice. Furthermore all 9 Lagrange multipliers $v_a^j$ of $T^a_j$ 
get fixed in that process. Finally, stabilisation of $T^a_j$ fixes 9 of 16
Lagrange multipliers $\hat{v}^A_I$ of $\hat{P}_A^I$. Next, the  
constraints $G_j, C, C_a$ are already stable while stabilisation of 
$D_j$ fixes further 3 of 16 of the $\hat{v}^A_I$. This ends the stabilisation 
process. 

All Lagrange multipliers but $4=16-9-3$ of the $\hat{v}^A_I$ and all
9 of the $v_a^j$ were fixed. The means that 4 of the $\hat{P}_A^I$, call 
them $\hat{P}_A$\ are 
first class (they are linear combinations of the momenta conjugate to lapse 
and shift functions). Furthermore, $\pi^t_j,G_j, C,C_a$ are first class while 
3 of the $\hat{P}_A^I$, call them $\hat{P}^j$ form second class pairs with 
$D_j$ while 9 of the  
$\hat{P}_A^I$, call them $\hat{P}_a^j$ form second class pairs with 
the $T^a_j$. Correspondingly we have $2 \cdot (3+9)=24$ second class constraints
$D_j, \hat{P}^j, T^a_j, \hat{P}_a^j$ and $3+3+4+4=14$ first class constraints
$\pi^t_j, G_j, C_a, C, \hat{P}_A$. These reduces $24+2 \cdot 14=52$ of the 
$2 \cdot (12+16)=56$ degrees of freedom $A_B^j, \pi^B_j, \hat{e}^A_I, \hat{P}_A^I$
leaving the 4 propagating degrees of freedom of the Hamiltonian U(1)$^3$ or 
SU(2) theory respectively.\\
\\
We now outline the details:\\
Plugging (\ref{3.26}) into (\ref{3.25}) results in the 3+1 decomposition
\be \label{3.27}
S=\int\; dt\; d^3x\;[F_{ta}^j\; \hat{\sigma}^{ta}_j+\frac{1}{2}\; F_{ab}^j\;
\hat{\sigma}^{ab}_j],\qquad
\hat{\sigma}^{AB}_j:=2\hat{\Sigma}^{AB}_{0j}+\gamma\epsilon^{jkl}\hat{\Sigma}^{AB}_{kl}
\ee
The computation of the conjugate momenta leads to the primary constraints
\be \label{3.28}
\pi^t_j=0,\qquad T^a_j=\pi^a_j-\hat{\sigma}^{ta}_j=0,\qquad  \hat{P}_A^I=0
\ee
and thus the primary Hamiltonian reads with the velocities $v_A^j,\hat{v}^A_I$
\ba \label{3.29}
H &=& \int\; d^3x\; [v_A^j \pi^A_j+\hat{v}^A_I \;\hat{P}_A^I-L]
\nonumber\\
&=&\int\; d^3x\; [v_t^j \pi^t_j+v_a^j \; T^a_j+\hat{v}^A_I \;\hat{P}_A^I
-A_t^j\; (\nabla_a \;\sigma^{ta}_j)
-\frac{1}{2}\; F_{ab}^j\;\sigma^{ab}_j]
\ea
where $\nabla$ denotes the U(1)$^3$ or SU(2) covariant derivative acting on 
Lie algebra indices only, that is, $\nabla_a T_j=\partial_a T_j$ and 
$\nabla_a T_j=\partial_a T_j+\epsilon_{jkl} A_a^k T_l$ respectively.  

Stabilisation of $\pi^t_j$ yields the Gauss constraint
\be \label{3.30}
G_j=\nabla_a \sigma^{ta}_j
\ee
Stabilisation of $\hat{P}_A^I$ yields the condition
\be \label{3.31}
\{H,\hat{P}_A^I(f^A_I)\} 
=\int\; d^3x\; \{\hat{P}_A^I(f^A_I),v_a^j \sigma^{ta}_j+\frac{1}{2} F_{ab}^j 
\hat{\sigma}^{ab}_j\}=0
\ee
for all $f^A_I$. Isolating the i. $A=t,\; I=0$, ii. $A=t,\; I=i$, iii.
$A=c,\; I=0$, iv. $A=c,\; I=i$ coefficients yields the following set of 
$1+3+3+9=16$ conditions
\ba \label{3.32}
0 &=& v_a^j \hat{e}^a_j
\nonumber\\
0 &=& v_a^i\; \hat{e}^a_0+\gamma\;\epsilon^{ikl}\; v_a^k\; \hat{e}^a_l
\nonumber\\
0 &=& -\hat{e}^t_j\; v^j_c+F^j_{cb}\; \hat{e}^b_j
\nonumber\\
0 &=& v_c^i \; \hat{e}^t_0+\gamma\; \epsilon^{ikl} \; v_c^k\; \hat{e}^t_l
-[F_{cb}^i\; \hat{e}^b_0+\gamma\; \epsilon^{ikl}\; F_{cb}^k\; \hat{e}^b_l]
\ea
The general solution of the system (\ref{3.32})
requires a detailed case by case analysis which is provided in the appendix \ref{appendix}.
\\
However, a physically motivated solution consists in the following 7 
secondary constraints 
\be \label{3.33}
D_j:=\hat{e}^t_j,\qquad C_a:=F_{ab}^j\; \hat{e}^b_j,\qquad 
C:=\epsilon_{jkl} F_{ab}^j\; \hat{e}^a_k\;\hat{e}^b_l
\ee
and the following restriction on $v_a^j$
\be \label{3.34}
v_a^j=\frac{1}{\hat{e}^t_0}\;[F_{ab}^j\; \hat{e}^b_0
+\gamma\; \epsilon^{jkl}\; F_{ab}^k\; \hat{e}^b_l]     
\ee
where we have assumed $\hat{e}^t_0\not=0$.
Indeed using $D_j=0$ in the fourth equation of (\ref{3.32}) results in 
(\ref{3.34}). Using $D_j=0$  in the third equation of (\ref{3.32}) results 
in $C_a=0$. Inserting (\ref{3.34}) into $\hat{e}^t_0$ times the first 
equation of (\ref{3.32}) yields $\hat{e}^a_0 C_a+\gamma\; C=0$ i.e. $C=0$.
Finally inserting (\ref{3.34}) into $\hat{e}^t_0$ 
times the second equation in (\ref{3.32})
yields  
\be \label{3.35}
0=\gamma^2\; \epsilon^{ikl}\; \epsilon^{kmn} \; F_{ab}^m \hat{e}^b_n \;
\hat{e}^a_l
=\gamma^2(-C_a\; \hat{e}^a_i - F_{ab}^i\; \hat{q}^{ab})
=-\gamma^2\; C_i
\ee
and is thus already satisfied. Here symmetry of $\hat{q}^{ab}=\delta^{jk}
\hat{e}^a_j\hat{e}^b_k$ was used.

Stabilisation of $T^a_j$ leads to
\be \label{3.36}    
(\nabla_b \hat{\sigma}^{ab}_j)+\hat{v}^t_0\hat{e}^a_j-\hat{v}^t_j \hat{e}^a_0
+\hat{e}^t_0\;\hat{v}^a_j+\epsilon^{jkl}\;\hat{v}^t_k\; \hat{e}^a_l=0
\ee
which can be solved for $\hat{v}^a_j$. 

Before doing so, we consider the stabilisation of the secondary constraints.
Obviously stabilisation of $D_j$ enforces 
\be \label{3.37}
\hat{v}^t_j=0
\ee
so that (3.36) simplifies to 
\be \label{3.36}    
(\nabla_b \hat{\sigma}^{ab}_j)+\hat{v}^t_0\hat{e}^a_j
+\hat{e}^t_0\;\hat{v}^a_j=0
\ee
The Gauss constraint can be substituted by $\hat{G}_j=\nabla_a \pi^a_j$ modulo
$T^a_j$ which is already stabilised. In this form it generates Gauss gauge 
transformations on the sector $(A_a^j,\pi^a_j)$. Since
\be \label{3.37}
F_{ab}^j \sigma^{ab}_j=2\hat{e}^a_0 C_a+C
\ee
depends only on invariants, it Poisson commutes with this part of $H$.
In the Abelian case it also Poisson commutes with all other parts 
of $H$ but in the non-Abelian case it does not Poisson commute with 
$T^a_j$. In the non-Abelian case 
we add terms to $\hat{G}_j$ linear in $\hat{P}_A^I$ which is 
already stabilised
\be \label{3.37a}
\hat{G}_j=\nabla_a \pi^a_j-\epsilon_{jkl}\;\hat{P}^A_k \;\hat{e}^A_l
\ee
and now $\hat{G}_j$ also generates SU(2) rotations in the 
$(\hat{e}^A_j,\hat{P}_A^j)$ sector while it leaves the sector 
$(\hat{e}^A_0,\hat{P}_A^0)$ invariant. In particular, it rotates 
$T^a_j$ into itself. It follows that $\hat{G}_j$ is stabilised and thus 
$G_j$ is stabilised also in the non-Abelian theory.

Next, modulo $T^a_j,G_j$ which are already stabilised 
we can substitute $C_a$ by 
$\hat{C}_a=F_{ab}^j\; \pi^b_j-A_a^j \nabla_b \pi^b_j$ 
in whose form it generates spatial 
diffeomorphisms on the variables $A_a^j,\pi^a_j$. We add terms linear 
in the already stabilised $\hat{P}_A^I$ so that it generates spatial 
diffeomorphisms on all variables (taking the half-density weight into 
account so that $\hat{e}^t_I$ and $\hat{e}^a_I$ are respectively scalar 
and vector half-densities)
\be \label{3.37b}
\hat{C}_a = F_{ab}^j\; \pi^b_j - A_a^j \nabla_b \pi^b_j
+\frac{1}{2}(\hat{P}_t^I \hat{e}^t_{I,a}
-\hat{P}_{t,a}^I \hat{e}^t_I+2(\hat{P}_a^I\hat{e}^b_I)_{,b}-
\hat{P}_{b,a}^I \hat{e}^b_I + \hat{P}_{b}^I \hat{e}^b_{I,a})
\ee
It follows that $\hat{C}_a$ and thus $C_a$ 
is already stabilised since H is a linear 
combination of constraints
and all constraints are tensor densities. As far as $C$ is concerned, we 
consider instead $\hat{C}$ which is the integrand of $H$ with 
the fixed expressions for $v_a^j,\; \hat{v}^A_j$. Then we compute
\be \label{3.38}
\{\hat{C}(f),\hat{C}(g)\}
=\frac{1}{2}\int\; d^3x\;d^3y\; [f(x)\; g(y)-f(y)\;g(x)]\;
\{\hat{C}(x),\hat{C}(y)\}
\ee
Modulo constraints, only the contributions to the Poisson bracket 
which lead to derivatives of the $\delta$ distribution do not vanish
in (\ref{3.38}). The only derivatives within constraints in $\hat{C}$ 
come from the term $-\frac{1}{2} F_{ab}^j \;\hat{\sigma}^{ab}_j$ which has 
non vanishing Poisson brackets leading to derivatives of 
$\delta$ distributions only with the term $v_a^j T^a_j$. It follows
\ba \label{3.39}
\{\hat{C}(f),\hat{C}(g)\}
&=& -\frac{1}{2}\int\; d^3x\;d^3y\; [f(x)\; g(y)-f(y)\;g(x)]\;v_a^j(x)
\{\pi^a_j(x),F_{bc}^k(y)\}\;\sigma^{bc}_k(y)
\nonumber\\
&=& -\int\; d^3x\;d^3y\; [f(x)\; g(y)-f(y)\;g(x)]\;v_a^j(x)
\{\pi^a_j(x),\partial_{[b}\; A_{c]}^k(y)\}\;\sigma^{bc}_k(y)
\nonumber\\
&=& \int\; d^3x\; [f\; g_{,b}-f_{,b}\;g]\;v_c^j\;\sigma^{bc}_jk(y)
\nonumber\\
&=& \int\; d^3x\;\omega_b\;
(F_{cd}^j+\gamma\epsilon^{jkl} F_{cd}^k\;\hat{e}^d_l)\;
(2\hat{e}^b_{[0}\;\hat{e}^c_{j]}+\epsilon^{jmn}\;\hat{e}^b_m\;\hat{e}^c_n)
\nonumber\\
&=& \int\; d^3x\;\omega_b\;
\{-C_d\;\hat{e}^d_0 \hat{e}^b_0-(F_{cd}^j \hat{e}^c_0
\hat{e}^d_0)\hat{e}^b_j-\gamma\; \hat{e}^b_0\; C
\nonumber\\
&&
+\gamma\;[\hat{e}^d_0\; F_{cd}^j\epsilon^{jmn}
\hat{e}^b_m\hat{e}^c_n- 
\hat{e}^c_0\; F_{cd}^k\epsilon^{jkl}
\hat{e}^d_l\hat{e}^b_j]+
\gamma^2\; F_{cd}^k\;\epsilon^{jmn}\epsilon^{jkl} \hat{e}^d_l
\hat{e}^b_m\hat{e}^c_n\}
\nonumber\\
&=&
\gamma\int\; d^3x\;\omega_b\;\{F_{cd}^k\;\hat{e}^b_m\;
[\hat{e}^d_0\; \epsilon^{kmn}\;\hat{e}^c_n- 
\hat{e}^c_0\; \epsilon^{mkn}\hat{e}^d_n]+
\gamma\; F_{cd}^k\;(\hat{e}^b_k\; \hat{q}^{cd}-\hat{e}^c_k\; \hat{q}^{bd})
\nonumber\\
&=&
\gamma\int\; d^3x\;\omega_b\;\{F_{cd}^k\;\hat{e}^b_m\;
[\hat{e}^d_0\; \epsilon^{kmn}\;\hat{e}^c_n- 
\hat{e}^c_0\; \epsilon^{mkn}\hat{e}^d_n]+
\gamma\; F_{cd}^k\;(\hat{e}^b_k\; \hat{q}^{cd}-\hat{e}^c_k\; \hat{q}^{bd})
\nonumber\\
&=&
\gamma\int\; d^3x\;\omega_b\;\{F_{cd}^k\;\hat{e}^b_m\;\epsilon^{kmn}\;
[\hat{e}^d_0\;\;\hat{e}^c_n+\hat{e}^c_0\;\hat{e}^d_n]+
\gamma\; \hat{q}^{bd}\; C_d\}
\nonumber\\
&=&
\gamma^2\int\; d^3x\;\omega_b\; \hat{q}^{bd}\; C_d
\ea
where in the first step we isolated the relevant terms, in the second we 
dropped the term in $F_{ab}^j$ quadratic in the connection as its Poisson 
bracket is ultra-local (this step is avoided in the Abelian case), in the 
third we carried out the Poisson bracket and integrated by parts, in the 
fourth we defined $\omega_b:=\frac{1}{\hat{e}^t_0}(f \; g_{,b}-f_{,b}\;g)$
to simplify the notation and explicitly made use of definition of 
$\hat{\sigma}^{ab}_j$ and (\ref{3.34}), in the fifth we explicitly wrote 
out the six terms arising from the multiplication of the round brackets in 
the previous step, in the sixth we dropped the terms proportional to 
$C, C_d$ and the term in round brackets which vanishes due to antisymmetry and 
relabelled indices in the square bracket term and used summation identities,
in the seventh we wrote out the square bracket term such that 
it is manifestly symmetric in $c,d$ and used the definition of $C_d$ and 
$F_{cd}^k \hat{q}^{cd}=0$ and in the final step we used antisymmetry of
$F_{cd}^k$. The identity (\ref{3.39}) is thus a manifestation of 
the hypersurface deformation algebra. Choosing $f=1$ shows that 
$\{H,\hat{C}(g)\}$ is weakly vanishing for all $g$ thus $\hat{C}$ is 
stabilised. Since $\hat{C}=C$ plus constraints that are already stabilised,
it follows that $C$ itself is also stabilised.

Accordingly there are no tertiary constraints and all secondary 
constraints are stabilised. All $v_a^j$ and all $\hat{v}^A_j$ have been 
fixed while $\hat{v}^A_t$ remain free. We come to the classification of the 
primary constraints $\pi^t_j,\; T^a_j,\; \hat{P}_A^I$ and secondary 
constraints $G_j, C, C_a, D_j$. \\
i. $\pi^t_j$\\
Since all constraints
are independent of $A_t^j$, $\pi^t_j$ is first class.\\
ii. $\hat{G}_j$\\
This constraint generates Gauss gauge transformations on 
$\pi^a_j, A_a^j, \hat{e}^A_j,\hat{P}_A^j$. All constraints either are 
invariant or covariant under Gauss transformations. Thus $\hat{G}_j$ is 
first class. \\
iii. $\hat{C}_a$\\
This constraint generates spatial diffeomorphisms on all variables and all
constraints are tensor densities. Hence $\hat{C}_a$ is first class.\\
iv. $\hat{C}$\\
Since $\hat{C}$ is the integrand of $H$ which stabilises all constraints,
it follows that $\hat{C}$ has weakly vanishing Poisson brackets with 
all constraints except possibly those whose Poisson brackets 
involve derivatives of $\delta$ distributions since $H=\hat{C}(f=1)$ 
rather than general $\hat{C}(f)$. 
These are the brackets with $\hat{C}(g)$ and with $T^a_j(g^a_j)$. The first 
bracket has been checked in (\ref{3.38}) and the second yields the same result
as with $H$ except that the integral of the resulting Poisson  bracket
also involves $f$ as an underived factor. It follows that $\hat{C}$ is 
first class.\\  
v. $\hat{P}_A^0$:\\
We only need to check its Poisson brackets with $\hat{P}_B^I, T^a_j, D_j$. Clearly
the brackets with $\hat{P}_B^I,\; D_j$ vanish exactly while 
\be \label{3.39}
\{\hat{P}_A^0(f^A),T_a^j(g^a_j)\}
=-\int\; d^3x\; f_A\; g^a_j\;
\frac{\partial \hat{\sigma}^{ta}_j}{\partial \hat{e}^A_0}  
=-\int\; d^3x\; g^a_j\;(f^0 \hat{e}^a_j-f^a D_j)
\ee
It follows that $\hat{P}_a^0$ is first class. We substitute $\hat{P}_t^0$
by $\hat{P}_t^{\prime 0}=\hat{P}_t^0-\frac{\hat{e}^a_j}{\hat{e}^t_0} 
\hat{P}_a^j$. Note that this quantity is Gauss invariant and a tensor density
hence we just need to check its Poisson brackets with $\hat{P}^B_I, D_j, T^a_I$. 
With $\hat{P}^B_I$ they vanish weakly and with $D_j$ exactly while 
\be \label{3.40}
\{\hat{P}_t^{\prime 0}(f),T_a^j(g^a_j)\}
=\gamma\int\; d^3x\; f\; \frac{\hat{e}^a_j}{\hat{e}^t_0}\; g^a_k\;
\epsilon_{jkl} D_l
\ee
It follows that $\hat{P}_t^{\prime 0}$ is also first class.\\
vi. $\hat{P}_t^j, \; D_k$:\\
These obviously form a second class pair
\be \label{3.41}
\{\hat{P}_t^j(x),D_k(y)\}=\delta_{jk} \delta(x,y)
\ee
vii. $T^a_j,\hat{P}_b^k$:\\
These also form a second class pair
\be \label{3.42}
\{\hat{T}_a^j(x),\hat{P}_b^k(y)\}
=\{\hat{P}_b^k(y),\sigma^{ta}_j(x)\}=\hat{e}^t_0\delta(x,y)\delta^a_b
\delta^k_j
\ee
modulo a term $\propto D_l$.\\
\\
Accordingly, we arrive precisely at the constraint structure as anticipated 
above. It remains to solve the second class constraints and to compute the 
Dirac bracket. To solve $D_j=0, P_A^j=0$ is trivial. Note that for $D_j=0$ 
we have $\sigma^{ta}_j=\hat{e}^t_0 \hat{e}^a_j$ thus solving 
$T^a_j=0$ is equivalent to $\pi^a_j=\hat{e}^t_0 \; \hat{e}^a_j$. 
Accordingly, we focus attention on the remaining variables 
$A_a^j, \pi^a_j, \hat{P}_A^0, \hat{e}^A_0$
and the 
spatial diffeomorphism and Hamiltonian constraint can equivalently be 
formulated as 
\be \label{3.43}
C_a=F_{ab}^j \; \pi^b_j,\qquad
C=F_{ab}^j \;\epsilon_{jkl} \pi^a_k\;\pi^b_l
\ee
Due to (\ref{3.41}) and (\ref{3.42}) the Dirac bracket between 
phase space functions $U,V$ is given by 
\be \label{3.44}
\{U,V\}^\ast=\{U,V\}
\pm(\int\;d^3x\;\{U,\pi_t^j(x)\}\;\{D_j(x),V\}  
-\; U\;\leftrightarrow\; V)
\pm(\int\;d^3x\;\frac{1}{\hat{e}^t_0(x)}\;
\{U,T^a_j(x)\}\;\{\hat{P}_a^j(x),V\}  
-\; U\;\leftrightarrow\; V)
\ee
If we restrict $U,V$ to be functions of  
$A_a^j, \pi^a_j, \hat{P}_A^0, \hat{e}^A_0$ then certainly 
$\{U,P_A^j\}=\{V,P_A^j\}=0$. Hence on those functions, the Dirac 
bracket coincides with the Poisson bracket. Finally, since we set the second 
class constraints strongly to zero we have 
\be \label{3.44}
\hat{G}_j=G_j=\nabla_a \pi^a_j, \qquad \hat{P}_t^{0\prime}=\hat{P}_t^0
\ee
and $H$ is a linear combination of $\hat{P}_A^0, C_a, C, G_j$. We note that 
when $D_j=0$ we have in terms of lapse and shift functions 
\be \label{3.45}
e^t_0=\frac{1}{N},\qquad e^a_0=-\frac{N^a}{N}
\ee
and $e^a_j$ is invertible. Then $\det(e_A^I)=N\det(e_a^i)$ and 
$\hat{P}_A^0$ are essentially the momenta conjugate to lapse and shift
(modulo a canonical transformation). Thus we arrive exactly at the 
Hamiltonian formulation of the U(1)$^3$ or SU(2) model (Euclidean GR)
respectively independent of the value of $\gamma\not=0$.

\section{Pure Connection Formulation}
\label{s4}

As has been discovered in tandem with the Ashtekar-Barbero variables,
there exist (almost) pure connection formulations for 
Lorentzian vacuum General Relativity.
Without cosmological constant, it is possible to construct a polynomial action 
in terms of a self-dual $SL(2,\mathbb{C})$ connection and a density 
weighted volume form
\cite{8}, whereas with the cosmological constant it is possible to even remove the 
volume form and arrive at a non-polynomial pure connection formulation
\cite{11}. In this section we revisit these considerations for Euclidean 
signature and arbitrary $\gamma$ parameter (twisted self-duality) and 
simultaneously for both U(1)$^3$ and SU(2). We follow closely \cite{11}
but are able to go one step further in the following sense:
In \cite{11} it was pointed out that imposing the Hamiltonian constraint 
as a primary constraint into the resulting action rather than a secondary 
constraint can be used {\it in principle} to arrive at a pure connection 
formulation. However, the equation to be solved was not written out in 
detail nor was it shown explicitly that it can be solved algebraically 
(it could be a polynomial equation in the Lagrange multiplier of 
higher than fourth order). In this section we show that the equation 
to be solved fortunately is just a quartic equation. We write it out 
explicitly in the form that can solved using the Cardano -- Ferrari set 
of formulae \cite{12}. The resulting Lagrangian is then a {\it spacetime 
diffeomorphism covariant and pure connection Lagrangian}
for a SU(2) (or the U(1)$^3$) gauge theory that is equivalent 
to Euclidean General Relativity (or its U(1)$^3$ truncation) with 
cosmological constant which is quite remarkable as it brings GR much 
closer in language to Yang-Mills theory and opens new possibilities 
for path integral formulations. The difference with Yang-Mills theory 
is that the 
Lagrangian of GR is non-polynomial in the connection.  
Note that all considerations of this section also apply 
to Lorentzian signature, however then the curvatures appearing are 
genuinely complex valued.\\
\\
Because the following calculations are involved, to avoid confusion, it is beneficial to state at the outset that the main result of this section is the achievement of the Lagrangian (\ref{4.32}). As we will see in detail, in the expression of (\ref{4.32}) the function $\hat{D}$ depends merely on the connection $A$ and $\hat{\omega}$ is the real solution of the equation (\ref{4.31}) all of whose coefficients are dependent only on $A$, so is $\hat{\omega}$ itself. Therefore, as a final result, the Lagrangian (\ref{4.32}) provides the pure connection formulation of the theory.\\
\\
We begin with the Hamiltonian of the previous section including a 
cosmological constant
\begin{align} \label{4.1}
&H=\int\; d^3x\;h,\;\;
h:=-A_t^j\;\nabla_a E^a_j+N^a C_a-\frac{\gamma}{2}\; N\; \bar{C}
-N\; \Lambda\; [\det(E)]^{1/2},\nonumber\\
&\qquad C_a=F_{ab}^j\; E^b_j,\qquad \bar{C}=\epsilon_{jkl} F_{ab}^j\; E^a_k \; E^b_l\;
[\det(E)]^{-1/2}
\end{align}
where we have written out $\hat{e}^A_0$ in terms of lapse and shift functions.
To derive an action purely in terms of $A_A^j$ we first perform the Legendre 
transform of (\ref{4.1}) with respect to the momentum $E^a_j$ conjugate 
to $A_a^j$. This still leaves us with an expression that depends on 
$N, N^a$. One then removes these by extremising that action with respect
to $N, N^a$ and resubstituting the respective solution into the action. 
As explained in \cite{11}, this leads to an action from which
$C_a=0$ and $C:=\frac{\gamma}{2}\; \bar{C}+\Lambda [\det(E)]^{1/2}=0$ follow as primary constraints when passing again to the Hamiltonian 
formulation, rather than as secondary constraints as we deduced in the 
previous section.   

The Legendre transform determines the velocity
\be \label{4.2}
\partial_t A_a^j
=\frac{\delta H}{\delta E^a_j}
=\nabla_a A_t^j+N^b\; F_{ba}^j-\tilde{N}\;\gamma \epsilon_{jkl} F_{ba}^k\; E^b_l
- \tilde{N}\;\frac{\Lambda}{2} \epsilon_{jkl}\; \epsilon_{abc}\; E^b_k\; E^c_l
\ee  
where we have defined the lapse $\tilde{N}=N\;[\det(E)]^{-1/2}$
of density weight $-1$ as an independent variable. By an abuse 
of notation we relabel $C [\det(E)]^{1/2}$ as $C$ again.
Equation (\ref{4.2}) 
can be rewritten as 
\be \label{4.3}
F_{ta}^j-N^b F_{ba}^j=N\; F^j_{na}
=-\tilde{N}\;\epsilon_{jkl}[
\gamma F_{ba}^k\; E^b_l
+\frac{\Lambda}{2} \epsilon_{abc}\; E^b_k\; E^c_l]
\ee
with the spacetime curvature $F_{AB}^j$ and the normal 
$n^t=\frac{1}{N},\;n^a=-\frac{N^a}{N}$. Assuming the magnetic field
$B^a_j:=\epsilon^{abc} F_{bc}^j/2$ to be non-degenerate we can decompose 
$E^a_j=B^a_k\; \psi^k_j$ for some matrix $\psi$ and can write (\ref{4.3})
in the equivalent form (assuming the spatial metric to be non-degenerate, 
we also have $\det(\psi)\not=0$) 
\be \label{4.4!}
-\frac{N \; F^j_{na}\; B^a_k}{\tilde{N}\det(B)}
=\gamma[{\rm Tr}(\psi)\;\delta^j_k-\psi^j_k]+\Lambda\; [[\psi^{-1}]^T]^j_k
\det(\psi)
\ee
To see which conditions are imposed on $\psi$ when $C_a=C=0$ hold we compute
\be \label{4.5}
C_a=\epsilon_{abc} B^c_l B^b_k \psi^k_j \delta^{lj}=0,\qquad
C=\det(B)\;\{\frac{\gamma}{2}\left( [{\rm Tr}(\psi)]^2-{\rm Tr}(\psi^2)\right)
+\Lambda\det(\psi)\}=0
\ee
Thus, raising and lowering the internal indices with the Kronecker 
$\delta$ it follows that $\psi=\psi^T$ and that 
${\rm Tr}(\psi^{-1})=-\Lambda\gamma^{-1}$.
We can use the antisymmetric part of $\psi_{jk}$ to remove the antisymmetric 
part of the r.h.s. of (\ref{4.4!}). Hence in what follows we take 
$\psi$ to be symmetric and thus (\ref{4.4!}) becomes 
\be \label{4.4}
-\frac{N \; F^{(j}_{na}\; B^a_l \; \delta^{k)l}}{\tilde{N}\det(B)}
=\gamma[{\rm Tr}(\psi)\;\delta^j_k-\psi^j_k]+\Lambda\; [\psi^{-1}]^j_k
\det(\psi)
\ee
The l.h.s. is related to a symmetric, 
covariant spacetime scalar density one weighted 
matrix 
\ba \label{4.5!}
\kappa^{jk}:=\frac{1}{4} \epsilon^{ABCD}\; F_{AB}^j\; F_{CD}^k
=F_{ta}^{(j} B^a_l \; \delta^{k)l}
=N \; F_{na}^{(j} B^a_l \; \delta^{k)l}
\ea 
and the scalar density of weight $-1$
\be \label{4.6}
w:=-\frac{1}{\tilde{N}\;\det(B)}
\ee
so that we find the density weight zero matrix identity
\be \label{4.7}
\Omega:=w \kappa=\gamma [{\rm Tr}(\psi) \; 1_3-\psi]+\Lambda\; \det(\psi)
\psi^{-1}
\ee
and the Lagrangian becomes 
\begin{align} \label{4.8}
L &= E^a_j\partial_t A_a^j-h\nonumber\\
&=
\tilde{N}\det(B)\{-{\rm Tr}(\Omega \psi)+\frac{\gamma}{2}[[{\rm Tr}(\psi)]^2
-{\rm Tr}(\psi^2)]+\Lambda\det(\psi)\}\nonumber\\
&=w^{-1}\{\frac{\gamma}{2}[[{\rm Tr}(\psi)]^2
-{\rm Tr}(\psi^2)]+2\Lambda\det(\psi)]\}
\end{align}
where in the second step we computed ${\rm Tr}(\Omega\psi)$ from (\ref{4.7}). 
We did not yet impose $C=0$ since we want to compare with the method followed
in \cite{11}, so we postpone this to a later stage.

In order to organise the subsequent straightforward but tedious computations 
we define the scalars 
\be \label{4.9}
T:={\rm Tr}(\Omega),\qquad
S:=T^2-{\rm Tr}(\Omega^2), \qquad
D:=\det(\Omega),\qquad
\tau:={\rm Tr}(\psi),\qquad 
\sigma:=\tau^2-{\rm Tr}(\psi^2), \qquad
\delta:=\det(\psi)
\ee
We will also need the Caley-Hamilton identity in three dimensions 
\be \label{4.10}
\psi^3=\delta\; 1_3-\frac{\sigma}{2}\; \psi+\tau\; \psi^2
\ee
which in the present symmetric case can also be verified by elementary 
means by passing to the diagonal form. 

The strategy of \cite{11} is i. to derive three relations between 
$T, S, D, \tau, \sigma, \delta$ from the master equation 
(\ref{4.7}) by taking traces of its powers, ii. to solve 
$\sigma,\delta$ in terms of $T,D,S$, iii. to insert the solution into 
the Lagrangian (\ref{4.8}), iv. to ask that the Lagrangian is 
stationary under variation of $w$ which determines $w$ in terms 
of $\kappa$ and v. to insert that solution into (\ref{4.8}). It is 
quite astonishing that one can get even to stage iii. since in stage ii. 
we obtain coupled algebraic equations of order three which when 
decoupling them may easily lead to polynomial equations of degree five 
or higher for which no algebraic solution can be found. Yet, it is 
possible to find an expression for $L$ in terms of $S, T, D$ in closed 
form. However, in stage iv. we encounter a {\it quartic} equation. While 
we still can provide a closed (Ferrari) formula for its solution (which 
however involves solving a cubic equation), its 
introduction in $L$, which itself depends non-polynomially and not 
even rationally on $w$, would fill several pages. 

Thus, after having performed all steps up to iv. in order 
to illustrate the 
arising algebraic complexity and because a derivation 
was not provided in \cite{11}, we return to stage i. and solve 
$C=0$ already at that level. Now there are three relations between 
$T,S,D$ and the two parameters $\tau, \delta$ because by (\ref{4.5}) 
$C=0$ is equivalent to $\sigma=\frac{2}{\gamma}\Lambda\delta$. This means that there is a constraint
among $S,T,D$ which leads to a polynomial in $w$ with $\kappa$ dependent
coefficients which is a depressed quartic equation. In this case, the Lagrangian  
depends rationally on $w$ so that the final solution is of a lower complexity.\\       
\\
{\it Stage i.}\\
The following computations are drastically simplified by rewriting (\ref{4.7})
in terms of the eigenvalues $\lambda_j, \mu_j$ of $\Omega, \psi$
respectively 
\be \label{4.11}
\lambda_j=\gamma \;[\tau-\mu_j]+\Lambda\;\frac{\delta}{\mu_j}
\ee
where 
\be \label{4.12}
T=\lambda_1+\lambda_2+\lambda_3,\qquad
S=2(
\lambda_1 \lambda_2+\lambda_2 \lambda_3+
\lambda_3 \lambda_1),\qquad
D=\lambda_1\lambda_2\lambda_3
\ee
and similar for $\tau, \sigma,\delta$.

Taking the trace of (\ref{4.7}) we find 
\be \label{4.13}
T=2\gamma \tau+\frac{\Lambda}{2}\sigma     
\ee
Next, multiplying out the r.h.s. of $S$ given in (\ref{4.12}),
we obtain after regrouping terms 
\be \label{4.14a}
\frac{S}{2}
=\gamma^2[\tau^2+\frac{\sigma}{2}]
+\Lambda\gamma[6\delta-\sigma\tau+\tau^3-{\rm Tr}(\psi^3)]
+\Lambda^2\delta\tau
\ee
Taking the trace of (\ref{4.10}) we can write the r.h.s. just in terms of 
$\tau,\sigma,\delta$
\be \label{4.14}
\frac{S}{2}
=\gamma^2[\tau^2+\frac{\sigma}{2}]
+\Lambda\gamma[3\delta+\frac{1}{2}\sigma\tau]
+\Lambda^2\delta\tau
\ee
Note that the r.h.s. is only a quadratic polynomial. If that was not the 
case, we would not be able to complete even step ii. 

Next, again multiplying out the r.h.s. of $D$ given in (\ref{4.12})
we obtain after regrouping terms 
\be \label{4.15a}
D=\gamma^3[-\delta+\frac{1}{2}\tau\sigma]+\gamma\Lambda^2\delta\sigma
+\Lambda^3\delta^2
+\gamma^2\Lambda[3\tau\delta+\frac{1}{2}\{[{\rm Tr}(\psi^2)]^2
-{\rm Tr}(\psi^4)\}]
\ee
Multiplying (\ref{4.10}) with $\psi$ and taking the trace we find 
\be \label{4.15b}
[{\rm Tr}(\psi^2)]^2-{\rm Tr}(\psi^4)  
 =\frac{1}{2}\sigma^2-4\delta\tau
\ee
so that (\ref{4.15a}) can again be written just in terms of 
$\tau,\sigma,\delta$
\be \label{4.15}
D=
\gamma^3[-\delta+\frac{1}{2}\tau\sigma]+\gamma\Lambda^2\delta\sigma
+\Lambda^3\delta^2
+\gamma^2\Lambda[\delta\tau+\frac{1}{4}\sigma^2]
\ee
Again it is remarkable that the r.h.s. of (\ref{4.16}) is only a 
quadratic polynomial.

Equations (\ref{4.13}), (\ref{4.14}) and (\ref{4.15}) in principle allow 
to express $\tau,\sigma,\delta$ in terms of $T, S, D$. However, the fact 
that these are a coupled system of one linear and two quadratic 
polynomials still could forbid a simple algebraic solution. To see 
this, imagine expressing $\tau$ in terms of $T,\sigma$, using (\ref{4.13}) 
in (\ref{4.14}) and (\ref{4.15}). Then we obtain a coupled system of two
quadratic equations in terms of $\sigma, \delta$. We can solve 
(\ref{4.14}) (which contains $\delta$ linearly) for $\delta$ in terms of 
$S, \sigma$ which is a fraction 
with a quadratic and linear polynomial in $\sigma$ in numerator
and denominator respectively. 
Substituting that solution into (\ref{4.15}) which contains 
the square of $\delta$ and after multiplying by the square of the 
denominator we obtain a quartic polynomial in $\sigma$ which in general
is very complicated to solve. 

Fortunately, these complications can be avoided because we are only 
interested in the combination of $\tau, \sigma$ that appears in  $L$.
First let $y:=\Lambda\delta+\frac{1}{2}\gamma\sigma$ then 
\be \label{4.16}
\frac{S}{2}=\gamma^2(\tau^2-\sigma)+3\gamma y+\Lambda\tau y,\;
D=-\gamma^3\delta+\Lambda y^2+\gamma^2\tau y
\ee
Next with $z=\Lambda y+\gamma^2 \tau$
\be \label{4.17}
\frac{S}{2}=\tau\; z +3\gamma\; y-\gamma^2\sigma,\qquad
D=-\gamma^3\delta+y\;z
\ee
whence
\be \label{4.18}
\Lambda D+\frac{\gamma^2}{2}S
=\gamma^3 \;l+z^2,\qquad l=\frac{1}{2}\gamma\sigma+2\Lambda \delta
\ee
Recalling (\ref{4.8}) and (\ref{4.13}), we can now observe that
\be \label{4.19}
L=w^{-1}\;l,\qquad z=\frac{1}{2}[\Lambda\; l+\gamma T]
\ee
Consequently, we obtain the quadratic equation
\be \label{4.20}
\Lambda D+\frac{\gamma^2}{4}[2S-T^2]
=[\gamma^3+\frac{1}{2}\gamma\Lambda T]\;l+\frac{\Lambda^2}{4} l^2
\ee
with the two solutions
\be \label{4.21}
l=-a\pm\sqrt{b+a^2},\qquad
a=\frac{2\gamma^3+\gamma\Lambda T}{\Lambda^2},\qquad
b=\frac{4}{\Lambda^2}\{
\Lambda D+\frac{\gamma^2}{4}[2S-T^2]\}
\ee
If we insist on a well defined $\Lambda\to 0$ limit, only the positive 
sign in front of the square root is allowed.

Equation (\ref{4.21}) is as far as \cite{11} went (modulo the fact that there
$\gamma=\pm i$ for Lorentzian (anti-)selfdual General Relativity). 
The possibility 
to remove the implicit appearance of $w$ within $L=w^{-1}\;l(w,\kappa)$ was 
mentioned in \cite{11} but not carried out because of the tremendous 
complexity of the resulting equations. To see what would be required 
we push this a little further and define the dimension-free
quantities ($F/\Lambda$ is dimensionless such as $N, E$)   
\be \label{4.22}
\hat{\kappa}=\frac{\kappa}{\Lambda^2},\qquad
\tilde{w}=\Lambda^3 w,\qquad  \hat{\Omega}=\Lambda\Omega=\tilde{w}\; \hat{\kappa}
\ee
and the corresponding dimension-free and $w$ independent quantities
\be \label{4.23}
\Lambda\; T=\hat{T}\; \tilde{w},\qquad
\Lambda^2\;S=\hat{S}\; [\tilde{w}]^2,\qquad
\Lambda^3\;D=\hat{D}\; [\tilde{w}]^3
\ee
Then, the Lagrangian can be rewritten as
\be \label{4.23}
L=w^{-1} \; l=[\tilde{w}]^{-1}\;\Lambda^3\; l
=\Lambda\;\gamma \hat{l},\qquad
\hat{l}=-\hat{a}\pm\sqrt{\hat{b}+\hat{a}^2},\qquad
\hat{a}=2\;\hat{w}^{-1}+\hat{T}
\hat{b}=4\hat{D}\hat{w}+2\hat{S}-\hat{T}^2,\qquad
\hat{w}=\frac{\tilde{w}}{\gamma^2}
\ee
The action (\ref{4.23}) is stationary with respect to $\hat{w}$ when 
\be \label{4.24}
L'=\gamma\Lambda[-\hat{a}'\pm\frac{\hat{b}'+2\hat{a}\hat{a}'}{2\hat{W}}]
=0,\qquad
\hat{W}=\sqrt{\hat{b}+\hat{a}^2}
\ee
where the prime denotes a derivative with respect to $\hat{w}$. This can 
be written as 
\be \label{4.25}
2\hat{a}'[\hat{W}\mp \hat{a}]=\pm \hat{b}'=2\hat{a}' 
\frac{\hat{b}}{\hat{W}\pm \hat{a}}
\ee
Solving for $\hat{W}$ we find 
\be \label{4.26}
\pm \hat{W}=\hat{a}+\frac{\hat{b}'}{2\hat{a}'}=
2\frac{\hat{a}'\hat{b}}{\hat{b}'}-\hat{a}
\ee
i.e.
\be \label{4.27}
\hat{a}'(\hat{a}'\hat{b}-\hat{a}\hat{b}')=[\frac{\hat{b}'}{2}]^2
\ee
Denoting $\alpha=4\hat{D},\;\beta=2\hat{S}-\hat{T}^2$, we finally obtain
\be \label{4.28}
\frac{2}{\hat{w}^2}[\frac{2}{\hat{w}^2}[\alpha\hat{w}+\beta]
+\alpha[2\hat{w}-1+\hat{T}]]=[\frac{\alpha}{2}]^2
\ee
After multiplying by $\hat{w}^4$ this becomes a quartic equation 
in depressed form (no cubic term) which can be solved by Ferrari's formula.
However, since this requires implicitly solving a non-depressed (the 
quadratic term is non-vanishing) cubic, the explicit formula for the four 
roots is not only very lengthy, but also its insertion into (\ref{4.23}) 
becomes extremely involved due to the appearance of the square root 
term.\\ 
\\
Above procedure imposes the Hamiltonian constraint as 
a secondary constraint by extremising with respect to 
the Langrange multiplier $w$. We obtain of course an equivalent result 
by imposing
$C=0$, that is, 
$\frac{\gamma\sigma}{2}+\Lambda\delta=0$, already in 
(\ref{4.13}), (\ref{4.14}) and (\ref{4.15}) which makes the equation to be 
solved more transparent. First we find the simplified system 
\ba \label{4.28}
D &=& -\gamma^3\;\delta
\nonumber\\
\frac{S}{2} &=& \gamma^2 \tau^2+2\gamma\Lambda \delta
\nonumber\\
T &=& 2\gamma \tau-\frac{\Lambda^2}{2\gamma}\delta 
\ea
Eliminating $\tau,\delta$ imposes the constraint on $T,S,D$ given by 
\be \label{4.29}
\frac{S}{2}+2\frac{\Lambda}{\gamma^2} \; D=\frac{1}{4}\;
[\frac{\Lambda^2}{\gamma^4} D-T]^2
\ee
which is equivalent to imposing $C=0$ i.e. choosing $\sigma$ appropriately.
In terms of the dimension free quantities 
\be \label{4.30}
\hat{w}=\Lambda^3\;w\;\gamma^{-2},\qquad
\hat{T}= T\; \Lambda^{-2}\; w^{-1},\qquad
\hat{S}= S\; \Lambda^{-4}\; w^{-2},\qquad
\hat{D}= D\; \Lambda^{-6}\; w^{-3},\qquad
\ee
we find 
\be \label{4.31}
\frac{\hat{S}}{2}+2\hat{w} \; \hat{D}=\frac{1}{4}\;
[\hat{w}^2\;\hat{D}-\hat{T}]^2
\ee
which is again a quartic equation in depressed form for $\hat{w}$. Note 
that the Lagrangian for $C=0$ becomes simply 
\be \label{4.32}
L=-w^{-1} \;\Lambda\; \gamma^{-3}\; D\; 
=-w^2 \;\Lambda^7\; \gamma^{-3}\; \hat{D}\; 
=-\hat{w}^2 \; \Lambda \gamma \; \hat{D}
\ee
which looks deceptively simple but of course the challenge lies in solving
the quartic equation (\ref{4.31}) for $\hat{w}$ in terms 
of $\hat{T},\hat{S},\hat{D}$ which are traces of polynomials 
in the matrix $\hat{\kappa}= \kappa\Lambda^{-2}$. We have access to the general solution of 
(\ref{4.31}) using the Cardano-Ferrari theory. We do not display it here 
explicitly because the formulas are quite lengthy. However, we want to know whether among the four roots there are real ones and if 
yes, under which conditions on $\kappa$. Examining its determinants, one concludes that the equation (\ref{4.31}) always has a real solution \textbf{unless} $\hat{S}<0$ and
\begin{align}
&\hat{T}<0, \; \; 0< \hat{D} < \frac{1}{54} \hat{T}(9 \hat{S} - 4 \hat{T}^2) + \frac{\sqrt{(2\hat{T}^2 - 3\hat{S})^3}}{27 \sqrt{2}} \nonumber\\
&\rm{OR}\nonumber \\
&\hat{T}>0, \; \; \frac{1}{54} \hat{T}(9 \hat{S} - 4 \hat{T}^2) - \frac{\sqrt{(2\hat{T}^2 - 3\hat{S})^3}}{27 \sqrt{2}} < \hat{D} <0.
\end{align} 
\\
The second task is to check that the Legendre transform of (\ref{4.32}) delivers $C_a=C=0$ as primary constraints.
In order to make arrangements for the ensuing calculations which are rather tedious, we define
\begin{align}
\mathfrak{A}:=&\; - \frac{\gamma \hat{\omega}}{\Lambda(2 + \hat{\omega}(\hat{T}- \hat{D} \hat{\omega}^2))}, \nonumber\\
\mathfrak{B}:=& \; \frac{\gamma}{6(2 + \hat{\omega}(\hat{T}- \hat{D} \hat{\omega}^2))},\nonumber\\
\mathfrak{M}^{ij} :=&\; \kappa^{ij} - \hat{\omega} \hat{D} (\kappa^{-1})^{ij},\nonumber\\
\mathfrak{N} :=& \;\hat{T}+ \hat{D} \hat{\omega}^2,\nonumber\\
\tilde{\mathfrak{N}}:=& \; 6 + 4 \hat{T} \hat{\omega} - 2 \hat{D} \hat{\omega}^3
\end{align}
\\ 
Some of the quantities associated with the matrix $\mathfrak{M}^{ij}$ that are needed below can be computed in terms of $\hat{S}, \hat{T}, \hat{D} , \hat{\omega}$
\begin{align}\label{quantities for M}
{\rm Tr}(\mathfrak{M})=& \; \hat{T}- \frac{\hat{S} \hat{\omega}}{2} \nonumber\\
{\rm Tr}(\mathfrak{M}^2)=& \; \hat{T}^2 - \hat{S} - 6 \hat{D} \hat{\omega} + \frac{\hat{S}^2 \hat{\omega}^2}{4}- 2 \hat{T} \hat{D} \hat{\omega}^2 \nonumber\\
[{\rm Tr}(\mathfrak{M})]^2 - {\rm Tr}(\mathfrak{M}^2) =&\;  \hat{S} + (6 \hat{D} - \hat{S} \hat{T}) \hat{\omega} + 2 \hat{T} \hat{D} \hat{\omega}^2 \nonumber\\
\det(\mathfrak{M})=& \; \hat{D} - \frac{1}{2}(\frac{\hat{S}^2}{2}-4 \hat{T} \hat{D}) \hat{\omega} + \hat{D} (\hat{T}^2 -\hat{S}) \hat{\omega}^2 - \hat{D}^2 \hat{\omega}^3
\end{align}
where in order to obtain the second equation, we used ${\rm Tr} (\hat{\kappa}^{-2})= \frac{1}{2 \hat{D}^2}\left(\frac{\hat{S}^2}{2}-4 \hat{T} \hat{D} \right)$ and for the last equation we made use of the following relation
\begin{equation}
\det (A+B) = \det (A) + \det (B) + \det (B) {\rm Tr}(A B^{-1}) + \det (A) {\rm Tr}(B A^{-1})
\end{equation} 
which is valid for all $3 \times 3$ invertible matrices $A, B$.
\\
\\
Starting from the Lagrangian $L=- \gamma \Lambda \hat{D} \hat{\omega}^2$, first we need to calculate the momentum conjugate to $A_a^j$. Since the Lagrangian depends on $\hat{\omega}$, the conjugate momentum depends on the variation of $\hat{\omega}$ with respect to $\dot{A}^j_a$. Taking variation of both sides of (\ref{4.31}) and isolating $\delta \hat{\omega}$, one finds
\begin{align}
\delta \hat{\omega} =&\frac{(\hat{T}- \hat{D} \hat{\omega}^2)(\delta \hat{T} - \hat{\omega}^2 \delta \hat{D})-4 \hat{\omega} \delta \hat{D} - \delta \hat{S}}{2\hat{D} (2 + \hat{\omega}(\hat{T}- \hat{D} \hat{\omega}^2))}\nonumber\\
=&
\frac{\hat{T}- \hat{D} \hat{\omega}^2}{2\hat{D} (2 + \hat{\omega}(\hat{T}- \hat{D} \hat{\omega}^2))}\delta \hat{T}
+\frac{-1}{2\hat{D} (2 + \hat{\omega}(\hat{T}- \hat{D} \hat{\omega}^2))}\delta \hat{S}
+\frac{-\hat{\omega}^2 (\hat{T}- \hat{D} \hat{\omega}^2) -4 \hat{\omega} }{2\hat{D} (2 + \hat{\omega}(\hat{T}- \hat{D} \hat{\omega}^2))}\delta \hat{D}
\end{align}
Thus, the momentum is obtained as
\begin{align}\label{Momentum}
\Pi^a_j := &\frac{\delta L}{\delta \dot{A}^j_a}= - \gamma \Lambda \frac{\delta (\hat{\omega}^2 \hat{D})}{\delta \dot{A}^j_a}= - \gamma \Lambda \left(2 \hat{D} \hat{\omega} \frac{\delta \hat{\omega}}{\delta \dot{A}^j_a}+ \hat{\omega}^2 \frac{\delta \hat{D}}{\delta \dot{A}^j_a}\right) \nonumber \\
=&
- \frac{\gamma \Lambda}{2  \hat{D} (2 + \hat{\omega}(\hat{T}- \hat{D} \hat{\omega}^2))} \left[2 \hat{D} \hat{\omega} (\hat{T}- \hat{D} \hat{\omega}^2)\frac{\delta \hat{T}}{\delta \dot{A}^j_a} -2 \hat{D} \hat{\omega} \frac{\delta \hat{S}}{\delta \dot{A}^j_a} -4 \hat{D} \hat{\omega}^2 \frac{\delta \hat{D}}{\delta \dot{A}^j_a}\right] \nonumber \\
=&
- \frac{\gamma \Lambda \hat{\omega}}{(2 + \hat{\omega}(\hat{T}- \hat{D} \hat{\omega}^2))} \left[(\hat{T}- \hat{D} \hat{\omega}^2)\frac{\delta \hat{T}}{\delta \dot{A}^j_a} -\frac{\delta \hat{S}}{\delta \dot{A}^j_a} -2 \hat{\omega} \frac{\delta \hat{D}}{\delta \dot{A}^j_a}\right]\nonumber \\
=&
- \frac{\gamma \hat{\omega}}{\Lambda(2 + \hat{\omega}(\hat{T}- \hat{D} \hat{\omega}^2))} \left[(\hat{T}- \hat{D} \hat{\omega}^2)B^a_j -(2 \hat{T} B^a_j -2  \kappa^{ij} B_i^{a}) -2 \hat{\omega} \hat{D} (\kappa^{-1})_{ji} B^a_i\right]\nonumber \\
=&
- \frac{\gamma \hat{\omega}}{\Lambda(2 + \hat{\omega}(\hat{T}- \hat{D} \hat{\omega}^2))} \left[2 \left(\kappa^{ij} - \hat{\omega} \hat{D} (\kappa^{-1})_{ji}\right) B^a_i-(\hat{T}+ \hat{D} \hat{\omega}^2)B^a_j \right]\nonumber \\
=&
\mathfrak{A} \left[2 \mathfrak{M}^{ij} B^a_i-\mathfrak{N} B^a_j \right]
\end{align}
where from the third to the fourth line, we have used the following variations

\begin{align}\label{Variationss}
\left(\frac{\delta \hat{\kappa} }{\delta \dot{A}_c^l }\right)^{ij}=&\; \Lambda^{-2} \delta_l^{(i} B^{j)c},\nonumber\\
\frac{\delta \hat{T}}{\delta \dot{A}_c^l}=& \; \Lambda^{-2} B^c_l,\nonumber\\
\frac{\delta \hat{S}}{\delta \dot{A}_c^l}=&\;  2 \hat{T} \frac{\delta \hat{T}}{\delta \dot{A}_c^l} - 2\hat{\kappa}^{ij} \left(\frac{\delta \hat{\kappa}}{\delta \dot{A}_c^l}\right)_{ij}= \Lambda^{-2}(2 \hat{T} B^c_l -2  \hat{\kappa}^{il} B_i^{c})\nonumber\\
\frac{\delta \hat{D}}{\delta \dot{A}_c^l} =& \; \hat{D} (\hat{\kappa}^{-1})_{ij} \left(\frac{\delta \hat{\kappa}}{\delta \dot{A}_c^l}\right)^{ij}= \Lambda^{-2} \hat{D} (\hat{\kappa}^{-1})_{ij} \delta_l^{(i} B^{j)c}= \Lambda^{-2} \hat{D} (\hat{\kappa}^{-1})_{lj} B^c_j
\end{align}
In the last line of (\ref{Variationss}), the Jacobi's formula $d(\det(M))=\det(M) {\rm Tr}(M^{-1} dM)$ has been used, which is valid for every invertible matrix $M$. 

As $F^i_{ab}= \epsilon_{cab}B^c_i$, from (\ref{Momentum}) we immediately conclude that
\begin{align}
C_a =& \; F^i_{ab} \Pi^b_i = \mathfrak{A} \epsilon_{cab}B^c_i  \left[2 \mathfrak{M}^{ji} B^b_j-\mathfrak{N} B^b_i \right] = 0
\end{align}
because the matrix $\mathfrak{M}$ is symmetric. This simply shows that the vector constraint is a primary constraint.
\\
Achieving a similar result for the Hamiltonian constraint requires more calculations.
\begin{align}\label{HC}
C=& \; \frac{\gamma}{2} \epsilon_{ijk} F^i_{ab} \Pi^a_j \Pi^b_k + \Lambda \det (\Pi) \nonumber \\
=& \;
\frac{\gamma}{2} \epsilon_{ijk} \epsilon_{c ab} B^c_i \Pi^a_j \Pi^b_k + \frac{\Lambda}{6} \epsilon_{ijk} \epsilon_{c ab} \Pi^c_i \Pi^a_j \Pi^b_k \nonumber \\
=& \;
\left( \frac{\gamma}{2} B^c_i  + \frac{\Lambda}{6} \Pi^c_i \right) \epsilon_{ijk} \epsilon_{c ab}  \Pi^a_j \Pi^b_k \nonumber\\
=& \;
\frac{\gamma}{2} \left(B^c_i  -  \frac{ \hat{\omega}}{3(2 + \hat{\omega}(\hat{T}- \hat{D} \hat{\omega}^2))} \left[2 \left(\kappa^{li} - \hat{\omega} \hat{D} (\kappa^{-1})_{il}\right) B^c_l-(\hat{T}+ \hat{D} \hat{\omega}^2)B^c_i \right] \right) \epsilon_{ijk} \epsilon_{c ab}  \Pi^a_j \Pi^b_k \nonumber\\
=& \;
\frac{\gamma}{6(2 + \hat{\omega}(\hat{T}- \hat{D} \hat{\omega}^2))} \left((6 + 4 \hat{T} \hat{\omega} - 2 \hat{D} \hat{\omega}^3)B^c_i  - 2 \hat{\omega}  \left(\kappa^{li} - \hat{\omega} \hat{D} (\kappa^{-1})_{il}\right) B^c_l \right) \epsilon_{ijk} \epsilon_{c ab}  \Pi^a_j \Pi^b_k \nonumber\\
=& \;
\mathfrak{B} \left(\tilde{\mathfrak{N}}B^c_i  - 2 \hat{\omega}  \mathfrak{M}^{il} B^c_l \right) \epsilon_{ijk} \epsilon_{c ab}  \Pi^a_j \Pi^b_k
\end{align}
\\
Using (\ref{Momentum}), we calculate $ \epsilon_{ijk} \epsilon_{c ab}  \Pi^a_j \Pi^b_k$ part of $C$ as
\begin{align}\label{LHC}
\epsilon_{ijk} \epsilon_{c ab}  \Pi^a_j \Pi^b_k =& \; \epsilon_{ijk} \epsilon_{c ab} \mathfrak{A}^2 \left[2 \mathfrak{M}^{mj} B^a_m-\mathfrak{N} B^a_j \right] \left[2 \mathfrak{M}^{nk} B^b_n-\mathfrak{N} B^b_k \right] \nonumber\\
=& \;
\epsilon_{ijk} \epsilon_{c ab} \mathfrak{A}^2 \left[4 \mathfrak{M}^{mj} \mathfrak{M}^{nk} B^b_n B^a_m -4  \mathfrak{M}^{mj}\mathfrak{N} B^b_k B^a_m +\mathfrak{N}^2 B^a_j B^b_k \right] \nonumber \\
=& \;
\epsilon_{ijk} \mathfrak{A}^2 \det(B) (B^{-1})^l_c \left[4 \mathfrak{M}^{mj} \mathfrak{M}^{nk} \epsilon_{lmn} -4  \mathfrak{M}^{mj}\mathfrak{N} \epsilon_{lmk} +\mathfrak{N}^2 \epsilon_{ljk}\right] \nonumber \\
=& \;
\mathfrak{A}^2 \det(B) (B^{-1})^l_c \left[4 \mathfrak{M}^{mj} \mathfrak{M}^{nk} \epsilon_{ijk} \epsilon_{lmn} -4  \mathfrak{M}^{mj}\mathfrak{N}(\delta^i_l \delta^j_m - \delta^i_m \delta^j_l) + 2\mathfrak{N}^2 \delta^i_l \right] \nonumber \\
=& \;
\mathfrak{A}^2 \det(B) (B^{-1})^l_c \left[4 \mathfrak{M}^{mj} \mathfrak{M}^{nk} \epsilon_{ijk} \epsilon_{lmn} +4 \mathfrak{M}^{il}\mathfrak{N} + (2\mathfrak{N}^2 -4\mathfrak{M}^j_j\mathfrak{N} )\delta^i_l \right]
\end{align}
Finally, plugging (\ref{LHC}) into (\ref{HC}) and simplifying, we have

\begin{align*}
C =& \mathfrak{B}\mathfrak{A}^2 \det(B) (B^{-1})^l_c  \left(\tilde{\mathfrak{N}}B^c_i  - 2 \hat{\omega}  \mathfrak{M}^{ir} B^c_r \right) \left[4 \mathfrak{M}^{mj} \mathfrak{M}^{nk} \epsilon_{ijk} \epsilon_{lmn} +4 \mathfrak{M}^{il}\mathfrak{N} + (2\mathfrak{N}^2 -4\mathfrak{M}^j_j\mathfrak{N} )\delta^i_l \right]\\
=&
\mathfrak{B}\mathfrak{A}^2 \det(B) \left(\tilde{\mathfrak{N}}\delta^l_i  - 2 \hat{\omega}  \mathfrak{M}^{il} \right) \left[4 \mathfrak{M}^{mj} \mathfrak{M}^{nk} \epsilon_{ijk} \epsilon_{lmn} +4 \mathfrak{M}^{il}\mathfrak{N} + (2\mathfrak{N}^2 -4\mathfrak{M}^j_j\mathfrak{N} )\delta^i_l \right]\\
=&
\mathfrak{B}\mathfrak{A}^2 \det(B)\\
&\left(\tilde{\mathfrak{N}}\left[4 \mathfrak{M}^{mj} \mathfrak{M}^{nk} \epsilon_{ijk} \epsilon_{imn} +4 \mathfrak{M}^i_i \mathfrak{N} + 3 (2\mathfrak{N}^2 -4\mathfrak{M}^j_j\mathfrak{N} ) \right]\right.\\
&
\left. \hspace{0.5 cm}- 2 \hat{\omega}  \left[4 \mathfrak{M}^{il} \mathfrak{M}^{mj} \mathfrak{M}^{nk} \epsilon_{ijk} \epsilon_{lmn} +4 \mathfrak{M}^{il} \mathfrak{M}^{il}\mathfrak{N} + (2\mathfrak{N}^2 -4\mathfrak{M}^j_j\mathfrak{N} )\mathfrak{M}^i_i \right]\right)\\
=&
\mathfrak{B}\mathfrak{A}^2 \det(B)\\
&\left(
\tilde{\mathfrak{N}}\left[4  ( [{\rm Tr}(\mathfrak{M})]^2 - {\rm Tr}(\mathfrak{M}^2)) -8\mathfrak{N} {\rm Tr}(\mathfrak{M}) + 6\mathfrak{N}^2 \right]
- 2 \hat{\omega}  \left[24 \det(\mathfrak{M}) +4 ({\rm Tr}(\mathfrak{M}^2)-[{\rm Tr}(\mathfrak{M})]^2 ) \mathfrak{N} + 2\mathfrak{N}^2 {\rm Tr}(\mathfrak{M})\right]\right)\\
=&
\mathfrak{B}\mathfrak{A}^2 \det(B)
\left(4 ([{\rm Tr}(\mathfrak{M})]^2 - {\rm Tr}(\mathfrak{M}^2))(\tilde{\mathfrak{N}}+ 2\hat{\omega}  \mathfrak{N})- (8\mathfrak{N}\tilde{\mathfrak{N}}+4 \hat{\omega} \mathfrak{N}^2 ){\rm Tr}(\mathfrak{M})+6 \tilde{\mathfrak{N}}\mathfrak{N}^2 - 48 \hat{\omega}  \det(\mathfrak{M})\right)
\end{align*}
\\
Now one can employ the relations (\ref{quantities for M}) and write the Hamiltonian constraint as a polynomial of $\hat{\omega}$
\begin{align}\label{Ply 7}
C=& \; \mathfrak{B}\mathfrak{A}^2 \det(B)\left[- 12 \hat{D}^3 \hat{\omega}^7 - 6 \hat{D}^2 \hat{S} \hat{\omega}^6 + 12 \hat{D}^2 \hat{T} \hat{\omega}^5 + 6 (14 \hat{D}^2 + 
    2 \hat{D} \hat{S} \hat{T}) \hat{\omega}^4 + 6 (12 \hat{D} \hat{S} + 2 \hat{D} \hat{T}^2) \hat{\omega}^3 \right. \nonumber\\
    &\left. \hspace{2.2cm}+ 6 (2 \hat{S}^2 + 20 \hat{D} \hat{T} - 
    \hat{S} \hat{T}^2 ) \hat{\omega}^2 + 6 (16 \hat{D} + 4 \hat{S} \hat{T} - 2 \hat{T}^3) \hat{\omega} + 24 \hat{S} - 12 \hat{T}^2 \right]
\end{align}
It can be concluded that $C$ arises as a primary constraint, if the equations (\ref{Ply 7}) and (\ref{4.31}) have a common real solution.
Although (\ref{Ply 7}) seems too complicated to be solved, not surprisingly one can write it as the product of two polynomials
\begin{equation}
C= \; -6 \mathfrak{B}\mathfrak{A}^2 \det(B) \left(2 \hat{D} \hat{\omega}^3 + \hat{S} \hat{\omega}^2 + 2 \hat{T} \hat{\omega} + 2 \right) \left(\hat{D}^2 \hat{\omega}^4 - 2 \hat{D} \hat{T} \hat{\omega}^2 - 8 \hat{D} \hat{\omega} + \hat{T}^2 - 2\hat{S} \right)=0
\end{equation}
which vanishes because $\hat{D}^2 \hat{\omega}^4 - 2 \hat{D} \hat{T} \hat{\omega}^2 - 8 \hat{D} \hat{\omega} + \hat{T}^2 - 2\hat{S}=0$ due to the equation (\ref{4.31}). This ends checking that $C=C_a =0$ arise as primary constraints.

\section{Conclusion and Outlook}
\label{s5} 

In this paper we have shown in a detailed analysis that the U(1)$^3$
model for canonical gravity has two Lagrangian formulations 
from which it derives: One is a Palatini-Holst type action which uses 
a tetrad, the other is a pure connection Lagrangian. Both Lagrangians
may serve as possible starting point for a path integral formulation and 
can be put into the language of spin foams (see 
\cite{13}
and references therein). As the gauge group
is Abelian, we expect that the resulting spin foam model can be much better 
controlled than their non-Abelian versions which thus could serve as 
an interesting test laboratory for the spin foam approach to LQG. 
Interestingly, our results immediately generalise from U(1)$^3$ to 
SU(2) so that one can also write a spin foam model for Euclidean General
Relativity but with gauge group SU(2) rather than SO(4) which may also 
lead to major simplifications even in the non-Abelian context. 
We leave this for future work.\\
\\
\\
{\bf Acknowledgements}\\
\\
S.B. thanks the Ministry of Science, Research and Technology of Iran and 
FAU Erlangen-N\"urnberg for financial support.

\newpage
\begin{appendix}

\section{The general solution of the system (\ref{3.32})}\label{appendix}
In this appendix we obtain the general solution of the system (\ref{3.32}). 
\\
\\
\textit{Case 1}: $\hat{e}^t_0 \neq 0$
\\
\\
First, we use the last equation of (\ref{3.32}) to obtain $v^i_a$. Since $\hat{e}^t_0 \neq 0$, one uses its inverse and gets
\begin{equation}\label{et, v}
v^i_a = \frac{1}{\hat{e}^t_0} (F^i_{ab}\hat{e}^b_0 + \gamma \epsilon^{ikl} F^k_{ab} \hat{e}^b_l - \gamma \epsilon^{ikl} v^k_a \hat{e}^t_l)
\end{equation}
This is not the desired solution because the r.h.s. still depends on $v^k_a$. To get rid of it, we just  
Plug (\ref{et, v}) into itself and simplify the expression
\begin{align*}
v^i_a =& \frac{1}{\hat{e}^t_0} \left(F^i_{ab}\hat{e}^b_0 + \gamma \epsilon^{ikl} F^k_{ab} \hat{e}^b_l - \frac{ \gamma \epsilon^{ikl}}{\hat{e}^t_0} (F^k_{ab}\hat{e}^b_0 + \gamma \epsilon^{kmn} F^m_{ab} \hat{e}^b_n - \gamma \epsilon^{kmn} v^m_a \hat{e}^t_n) \hat{e}^t_l \right)\\
=&
\frac{1}{\hat{e}^t_0} \left(F^i_{ab}\hat{e}^b_0 + \gamma \epsilon^{ikl} F^k_{ab} \hat{e}^b_l - \frac{ \gamma \epsilon^{ikl}}{\hat{e}^t_0} (F^k_{ab}\hat{e}^b_0 + \gamma \epsilon^{kmn} F^m_{ab} \hat{e}^b_n)\hat{e}^t_l +\frac{ \gamma^2 }{\hat{e}^t_0} (\delta^i_n \delta^l_m - \delta^i_m \delta^l_n) v^m_a \hat{e}^t_n \hat{e}^t_l \right)\\
=&
\frac{1}{\hat{e}^t_0} \left(F^i_{ab}\hat{e}^b_0 + \gamma \epsilon^{ikl} F^k_{ab} \hat{e}^b_l - \frac{ \gamma \epsilon^{ikl}}{\hat{e}^t_0} (F^k_{ab}\hat{e}^b_0 + \gamma \epsilon^{kmn} F^m_{ab} \hat{e}^b_n)\hat{e}^t_l +\frac{ \gamma^2 }{\hat{e}^t_0} ([v^l_a \hat{e}^t_l] \hat{e}^t_i  - v^i_a \hat{e}^t_l \hat{e}^t_l)  \right)\\
=&
\frac{1}{\hat{e}^t_0} \left(F^i_{ab}\hat{e}^b_0 + \gamma \epsilon^{ikl} F^k_{ab} \hat{e}^b_l - \frac{ \gamma \epsilon^{ikl}}{\hat{e}^t_0} (F^k_{ab}\hat{e}^b_0 + \gamma \epsilon^{kmn} F^m_{ab} \hat{e}^b_n)\hat{e}^t_l +\frac{ \gamma^2 }{\hat{e}^t_0} (F^l_{ab} \hat{e}^b_l \hat{e}^t_i  - v^i_a \hat{e}^t_l \hat{e}^t_l)  \right)
\end{align*}
where in the last step, we have used the third equation of (\ref{3.32}). Now by moving the last term of the r.h.s. to the l.h.s., 
one can isolate $v^i_a$ and obtain  
\begin{align}\label{v}
v^i_a &= \frac{\hat{e}^t_0}{(\hat{e}^t_0 \hat{e}^t_0 + \gamma^2 \hat{e}^t_l \hat{e}^t_l)} \left(F^i_{ab}\hat{e}^b_0 + \gamma \epsilon^{ikl} F^k_{ab} \hat{e}^b_l - \frac{ \gamma \epsilon^{ikl}}{\hat{e}^t_0} (F^k_{ab}\hat{e}^b_0 + \gamma \epsilon^{kmn} F^m_{ab} \hat{e}^b_n)\hat{e}^t_l +\frac{ \gamma^2 }{\hat{e}^t_0} F^l_{ab} \hat{e}^b_l \hat{e}^t_i   \right)\nonumber\\
&=
\frac{\hat{e}^t_0}{(\hat{e}^t_0 \hat{e}^t_0 + \gamma^2 \hat{e}^t_l \hat{e}^t_l)} \left(F^i_{ab}\hat{e}^b_0 + \gamma \epsilon^{ikl} F^k_{ab} \hat{e}^b_l - \frac{ \gamma \epsilon^{ikl}}{\hat{e}^t_0} F^k_{ab}\hat{e}^b_0 \hat{e}^t_l + \frac{ \gamma^2 }{\hat{e}^t_0}( F^i_{ab} \hat{e}^b_j \hat{e}^t_j - 2 F^j_{ab} \hat{e}^b_{[i} \hat{e}^t_{j]}) \right)
\end{align}
With the assumption $e^t_j = 0$, (\ref{v}) reduces to (\ref{3.34}). Now by substituting (\ref{v}) in the second, third and first equations of (\ref{3.32}), respectively, the following constraints arise
\begin{align}\label{constraints1}
\tilde{C}_a &:=
 \left(1- \frac{\gamma^2 \hat{e}^t_i \hat{e}^t_i}{\hat{e}^t_0\hat{e}^t_0 + \gamma^2 \hat{e}^t_k \hat{e}^t_k} \right) F^l_{ab}\hat{e}^b_l - \frac{\hat{e}^t_i \hat{e}^t_0}{\hat{e}^t_0 \hat{e}^t_0 + \gamma^2 \hat{e}^t_k \hat{e}^t_k} (F^i_{ab} \hat{e}^b_0 + \gamma \epsilon^{ikl}F^k_{ab}\hat{e}^b_l )\nonumber\\
\tilde{C}_i &:=
\frac{\gamma^2 \hat{e}^t_0}{(\hat{e}^t_0 \hat{e}^t_0 + \gamma^2 \hat{e}^t_l \hat{e}^t_l)}  \left( \frac{\hat{e}^a_0}{\hat{e}^t_0}F^l_{ab} ( 2 \hat{e}^b_l \hat{e}^t_i - \hat{e}^b_i \hat{e}^t_l)+ F^{l}_{ab} \hat{e}^a_l \hat{e}^{b}_i 
+\frac{ \gamma}{\hat{e}^t_0}\epsilon^{ikl}\hat{e}^a_l( F^k_{ab} \hat{e}^b_j \hat{e}^t_j - 2 F^j_{ab} \hat{e}^b_{[k} \hat{e}^t_{j]}) \right)\nonumber\\
\tilde{C} &:= 
\frac{\hat{e}^t_0}{(\hat{e}^t_0 \hat{e}^t_0 + \gamma^2 \hat{e}^t_l \hat{e}^t_l)} \left(F^i_{ab}\hat{e}^b_0 \hat{e}^a_i + \gamma \epsilon^{ikl} F^k_{ab} \hat{e}^b_l \hat{e}^a_i - \frac{ \gamma \epsilon^{ikl}}{\hat{e}^t_0} F^k_{ab}\hat{e}^b_0 \hat{e}^t_l \hat{e}^a_i \right) 
\end{align}
We would like to check whether under the assumption of $\hat{e}^t_j = 0$ the above constraints (\ref{constraints1}) reduce to (\ref{3.33}). We begin with $\tilde{C}_a = 0$. It is obvious that putting $\hat{e}^t_j = 0$ in this equation yields to $F^l_{ab}\hat{e}^b_l =0$ which is the second constraint of (\ref{3.33}), i.e. $C_a=0$.\\
As most of the terms in $\tilde{C}_i$ are proportional to $\hat{e}^t_j $, when the latter equals to zero the constraint $\tilde{C_i}=0$ reduces to the simple equation $0=F^{l}_{ab} \hat{e}^a_l \hat{e}^{b}_i = -C_b \hat{e}^{b}_i$ that is not an independent constraint because we have already $C_a = 0$. Note that when we assume $\hat{e}^t_j=0$, we are, in fact, substituting three constraints $D_j=0$ for the above constraint $\tilde{C}_j=0$ and that is why in this case $\tilde{C}_j=0$ do not give us new independent constraints.\\
Finally, by putting $\hat{e}^t_j = 0$ in the equation $\tilde{C}=0$, we come to the conclusion that $0= F^i_{ab}\hat{e}^b_0 \hat{e}^a_i + \gamma \epsilon^{ikl} F^k_{ab} \hat{e}^b_l \hat{e}^a_i = -C_b\;\hat{e}^b_0 + \gamma \epsilon^{ikl} F^k_{ab} \hat{e}^b_l \hat{e}^a_i$ whose first term already vanishes. Therefore, $ \epsilon^{ikl} F^k_{ab} \hat{e}^b_l \hat{e}^a_i=0$ serves as an independent constraint which is the third constraint in (\ref{3.33}), i.e. $C=0$.
\\
\\
\textit{Case 2}: $\hat{e}^t_0 = 0$ 
\\
\\
In this case, in order to prevent degeneracy, at least one of $\hat{e}^t_i$ must be non-zero. Without loss of generality we can assume that $\hat{e}^t_1 \neq 0$. A consequence of this choice is separation of quantities with index $i=1$ and those with $i=\alpha\in \{2,3\}$. Greek alphabet is used to indicate indices $i\ne 1$.\\
The last equation of (\ref{3.32}) is employed to obtain $v^\alpha_a$ as   

\begin{equation}\label{A.4}
v^\alpha_a = \frac{1}{\gamma \hat{e}^t_1}(\gamma v^1_a \hat{e}^t_\alpha - \epsilon_{\alpha \beta} [F^\beta_{ab} \hat{e}^b_0+ \gamma \epsilon^{\beta k l}F^k_{ab}\hat{e}^b_l])
\end{equation}
and the third equation can be solved for $v^1_a$
\begin{equation}\label{A.5}
v^1_a = \frac{1}{\hat{e}^t_1}(F^j_{ab} \hat{e}^b_j - \hat{e}^t_\alpha v^\alpha_a)
\end{equation}
Now inserting (\ref{A.4}) in (\ref{A.5}) gives $v^1_a$ completely in terms of the canonical variables
\begin{equation}\label{A.6}
v^1_a  = \frac{\hat{e}^t_1}{\hat{e}^t_i \hat{e}^t_i}\left(F^j_{ab} \hat{e}^b_j + \hat{e}^t_\alpha \frac{1}{\gamma \hat{e}^t_1} \epsilon_{\alpha \beta} [F^\beta_{ab} \hat{e}^b_0+ \gamma \epsilon^{\beta k l}F^k_{ab}\hat{e}^b_l]\right)
\end{equation}
and plugging (\ref{A.6}) in (\ref{A.4}) finishes the restrictions on Lagrange multipliers $v^i_a$, as
\begin{equation}\label{A.7}
v^\alpha_a = \frac{\hat{e}^t_\alpha}{\hat{e}^t_i \hat{e}^t_i}\left(F^j_{ab} \hat{e}^b_j + \hat{e}^t_\delta \frac{\epsilon_{\delta \beta}}{\gamma \hat{e}^t_1} [F^\beta_{ab} \hat{e}^b_0+ \gamma \epsilon^{\beta k l}F^k_{ab}\hat{e}^b_l]\right) 
- \frac{\epsilon_{\alpha \beta}}{\gamma \hat{e}^t_1} [F^\beta_{ab} \hat{e}^b_0+ \gamma \epsilon^{\beta k l}F^k_{ab}\hat{e}^b_l])
\end{equation}
What is left to do is just substitution of $v^i_a$ into the rest of the equation of (\ref{3.32}) to attain the constraints. These equations are of the form
\begin{align}
0 =& v^1_a \hat{e}^a_0 + \gamma \epsilon^{\alpha \beta} v^\alpha_a \hat{e}^a_\beta\nonumber\\
0 =& v^\alpha_a \hat{e}^a_0 - \gamma \epsilon^{\alpha \beta} v^1_a \hat{e}^a_\beta + \gamma \epsilon^{\alpha \beta } v^\beta_a \hat{e}^a_1\nonumber\\
0 = & \hat{e}^a_1 v^1_a + \hat{e}^a_\alpha v^\alpha_a
\end{align}
where the first one is just the $i=1$ component of the last equation of (\ref{3.32}). After simplifying, the constraints acquire the following forms
\begin{align}
\tilde{C}_a :=&\; -\frac{\hat{e}^t_\alpha}{\hat{e}^t_1} [F^\alpha_{ab} \hat{e}^b_0 + \gamma \epsilon^{\alpha kl}F^k_{ab}\hat{e}^b_l]\nonumber\\
\tilde{C}_1 :=&\; \frac{\hat{e}^t_1}{\hat{e}^t_i \hat{e}^t_i}\left(F^j_{ab} \hat{e}^b_j + \hat{e}^t_\alpha \frac{1}{\gamma \hat{e}^t_1} \epsilon_{\alpha \beta} [F^\beta_{ab} \hat{e}^b_0+ \gamma \epsilon^{\beta k l}F^k_{ab}\hat{e}^b_l]\right) \hat{e}^a_0 
+ \gamma \epsilon^{\alpha \beta} v^\alpha_a \hat{e}^a_\beta \nonumber\\
\tilde{C}_\alpha :=& \;
\frac{\hat{e}^a_0 \hat{e}^t_\alpha}{\hat{e}^t_i \hat{e}^t_i}\left(F^j_{ab} \hat{e}^b_j + \hat{e}^t_\delta \frac{\epsilon_{\delta \beta}}{\gamma \hat{e}^t_1} [F^\beta_{ab} \hat{e}^b_0+ \gamma \epsilon^{\beta k l}F^k_{ab}\hat{e}^b_l]\right) 
- \frac{\epsilon_{\alpha \beta}}{\gamma \hat{e}^t_1} [F^\beta_{ab} \hat{e}^b_0+ \gamma \epsilon^{\beta k l}F^k_{ab}\hat{e}^b_l] \hat{e}^a_0 \nonumber\\ 
&-
\gamma \epsilon^{\alpha \beta} \hat{e}^a_\beta \frac{\hat{e}^t_1}{\hat{e}^t_i \hat{e}^t_i}\left(F^j_{ab} \hat{e}^b_j + \hat{e}^t_\delta \frac{1}{\gamma \hat{e}^t_1} \epsilon_{\delta \sigma} [F^\sigma_{ab} \hat{e}^b_0+ \gamma \epsilon^{\sigma k l}F^k_{ab}\hat{e}^b_l]\right)\nonumber \\
&+
\gamma \epsilon^{\alpha \beta } \hat{e}^a_1 \frac{\hat{e}^t_\beta}{\hat{e}^t_i \hat{e}^t_i}\left(F^j_{ab} \hat{e}^b_j + \hat{e}^t_\delta \frac{\epsilon_{\delta \sigma}}{\gamma \hat{e}^t_1} [F^\sigma_{ab} \hat{e}^b_0 + \gamma \epsilon^{\sigma k l}F^k_{ab}\hat{e}^b_l]\right) 
- \gamma \epsilon^{\alpha \beta } \hat{e}^a_1 \frac{\epsilon_{\beta \sigma}}{\gamma \hat{e}^t_1} [F^\sigma_{ab} \hat{e}^b_0+ \gamma \epsilon^{\sigma k l}F^k_{ab}\hat{e}^b_l]))\nonumber\\
\tilde{C}:= &\;
 \hat{e}^a_1 \frac{\hat{e}^t_1}{\hat{e}^t_i \hat{e}^t_i}\left(F^j_{ab} \hat{e}^b_j + \hat{e}^t_\alpha \frac{1}{\gamma \hat{e}^t_1} \epsilon_{\alpha \beta} [F^\beta_{ab} \hat{e}^b_0+ \gamma \epsilon^{\beta k l}F^k_{ab}\hat{e}^b_l]\right)\nonumber\\
&+
\hat{e}^a_\alpha \frac{\hat{e}^t_\alpha}{\hat{e}^t_i \hat{e}^t_i}\left(F^j_{ab} \hat{e}^b_j + \hat{e}^t_\delta \frac{\epsilon_{\delta \beta}}{\gamma \hat{e}^t_1} [F^\beta_{ab} \hat{e}^b_0+ \gamma \epsilon^{\beta k l}F^k_{ab}\hat{e}^b_l]\right) 
-\hat{e}^a_\alpha  \frac{\epsilon_{\alpha \beta}}{\gamma \hat{e}^t_1} [F^\beta_{ab} \hat{e}^b_0+ \gamma \epsilon^{\beta k l}F^k_{ab}\hat{e}^b_l]
\end{align}
To summarise, solving 16 equations (\ref{3.32}) to ensure that $\hat{P}^I_A$ is stabilised leads to fixing 9 Lagrange multipliers $v_a^1$ and $v_a^\alpha$ through (\ref{A.6}) and (\ref{A.7}), respectively, in addition to 7 secondary constraints $\tilde{C}_a,\tilde{C}_1,\tilde{C}_\alpha,\tilde{C}$.  
This ends the analysis.

\section{Hamiltonian analysis of the action of \cite{0a}}\label{appendix2}
In this appendix, we discuss the Hamiltonian analysis of the action introduced in \cite{0a}, that is
\begin{equation}\label{Initial action}
S= \frac{1}{2}\int d^4x\; \epsilon^{\mu \nu \alpha \beta} e^i_\mu e^j_\nu F_{\alpha \beta}^{ij}
\end{equation}
where $\mu, \nu, \cdots \in \{0, 1, 2, 3\}$ are spacetime indices and $i, j , \cdots \in {1, 2, 3}$ are Lie algebra indices.\\
The 3+1 decomposition of \ref{Initial action} is given by
\begin{align}\label{Smolin action}
S &=\int dt \; \int d^3x \; \epsilon^{abc} \left(e^i_t e^j_a F_{bc}^{ij} + e^i_a e^j_b F_{tc}^{ij} \right)
\end{align}
The conjugate momenta corresponding the configuration variables can easily be computed as
\begin{equation}
P_i:= \frac{\delta S}{\delta \dot{e}^i_t}=0, \;\; P^a_i:= \frac{\delta S}{\delta \dot{e}^i_a}=0, \;\; \pi_{ij}:= \frac{\delta S}{\delta \dot{A}^{ij}_t}=0, \;\; \pi^a_{ij}:= \frac{\delta S}{\delta \dot{A}^{ij}_a}=\epsilon^{abc} e^i_b e^j_c
\end{equation}
that lead to the primary constraints
\begin{equation}\label{Primary C S}
P_i=0, \;\; P_i^a=0, \;\; \pi_{ij}=0, \;\; T^a_{ij}:=\pi^a_{ij}-\epsilon^{abc} e^i_b e^j_c=0
\end{equation}
The Legendre transform of (\ref{Primary C S}) yields 
\begin{align}
H &= \int d^3x \; \{v^i P_i + v^i_a P^a_i + v^{ij}\pi_{ij} +  v_a^{ij} \pi^a_{ij} - L\}\\
&=
\int d^3x \; \{v^i P_i + v^i_a P^a_i + v^{ij}\pi_{ij} +  v_a^{ij} T^a_{ij} - \epsilon^{abc} e^i_t e^j_a F_{bc}^{ij} + \epsilon^{abc} e_a^i e_b^j  A_{t,c}^{ij}\}
\end{align}
where $v^i, v^i_a, v^{ij}, v_a^{ij}$ are the Lagrange multipliers and $L$ is the Lagrangian associated with the action (\ref{Smolin action}).\\
Stability of $\pi_{ij}=0$ leads to the Gauss secondary constraint
\begin{equation}
G_{ij}= \epsilon^{abc}\partial_c (e^i_a e^j_b) 
\end{equation}
Stability of $P_i=0$ yields 3 secondary constraints
\begin{equation}
C_i = \epsilon^{abc} e^j_a   F_{bc}^{ij}
\end{equation}
Stability of $P_i^a=0$ is obtained if and only if the following 9 equations are satisfied
\begin{align}
0 =&\; 2(A^{ij}_{t,c} - v^{ij}_c) \epsilon^{cab} e^j_b - \epsilon^{abc} e^j_t F^{ji}_{bc}\nonumber \\
:=&
\; f^{ij}_c \epsilon^{cab} e^j_b - \epsilon^{abc} e^j_t F^{ji}_{bc}\nonumber\\
=&
\; f^{ij}_k e^k_c \epsilon^{cab} e^j_b - \epsilon^{abc} e^j_t F^{ji}_{bc} \nonumber\\
=&
\; f_{ijk}  \epsilon^{klj} e^a_l \det(e) - \epsilon^{abc} e^j_t F^{ji}_{bc}
\end{align}
that are equivalent to
\begin{equation}\label{2}
f_{ijk}\epsilon^{klj} = \epsilon^{abc} e^m_t F^{mi}_{bc} e_a^l\det(e)^{-1} =: M_i^l
\end{equation}
Since $f_{ijk}= 2(A_{t,c}^{ij}-v_c^{ij})e^c_k$ is antisymmetric in $i,j$ we may
write it as
$f_{ijk}=\epsilon_{ijm} g^m_k$ for some matrix $g^m_k$.
Thus (\ref{2}) can be rewritten as
\begin{equation}\label{Original equation}
g^m_m \delta_i^l-g^l_i=M_i^l
\end{equation}
whose trace amounts to $2 g^m_m = M_m^m$. Plugging this into (\ref{Original equation}), we get
\begin{equation}
g^l_i = \frac{1}{2}M_m^m\delta_i^l -M_i^l
\end{equation}
meaning that all of the $f_{ijk}$ are fixed and $P_i^a=0$ is stabilised.
\\
Stability of $T^a_{ij}=0$ yields the equation
\begin{equation}\label{3}
\epsilon^{cba} \partial_b \left(e^{[i}_t e^{j]}_c \right) + \epsilon^{abc} v^{[i}_b e^{j]}_c =0
\end{equation}
Defining $v^b_a := v^i_a e_i^b$, we observe
\begin{align}\label{MR}
 \epsilon^{abc} v^d_b e^{[i}_d e^{j]}_c = -\epsilon^{cba} \partial_b \left(e^{[i}_t e^{j]}_c \right) =: M^a_{ij}
\end{align}
Multiplying both sides in $e_{[i}^e e_{j]}^f$ results in
\begin{equation}\label{Fi}
\epsilon^{ab[f} v^{e]}_b  = M^a_{ij} e_{[i}^e e_{j]}^f 
\end{equation}
and again by multiplying it in $\epsilon_{acd}$ we get
\begin{equation}
2\delta^{[f}_d v^{e]}_c =\epsilon_{acd}  M^a_{ij} e_{[i}^e e_{j]}^f
\end{equation}
from which the solution follows as
\begin{equation}
v^d_b = \frac{1}{2} M^a_{ij} \epsilon_{abc} e^d_{[i} e^c_{j]}
\end{equation}
Hence by fixing all the $v^i_a$ the stability of $T^a_{ij}=0$ is obtained. 
\\
Therefore, all the primary constraints are stabilised. Now, we want to check stability of the secondary ones.\\
\\
Stabilisation of $G_{ij}$ is already achieved, since $T^a_{ij}$ and $\pi^c_{ij,c}$ are stable. Modulo $T^a_{ij}=0$ the Gauss constraint is equivalent to $\hat{G}_{ij}= \pi^c_{ij,c}$ generating U(1)$^3$ gauge transformations.\\
\\
To check the stability of $C_i$, first note that it is equivalent to $C_a:= e^i_a C_i= \epsilon^{dbc} e^j_d  e^i_a  F_{bc}^{ij}= \epsilon^{dbc} e^j_{[d}  e^i_{a]}  F_{bc}^{ij}$. Looking at (\ref{Primary C S}), one immediately conclude that, modulo $T^a_{ij}=0$, the equation $\epsilon_{abc} \pi^a_{ij} = 2 e^i_{[b} e^j_{c]}$ holds. Therefore, modulo $T^a_{ij}=0$, we have
\begin{equation}
C_a = \epsilon^{dbc} e^j_{[d}  e^i_{a]}  F_{bc}^{ij} = \frac{1}{2} \epsilon^{dbc} \epsilon_{eda} \pi^e_{ij}F_{bc}^{ij} = -\frac{1}{2} (\delta^b_e \delta^c_a - \delta^b_a \delta^c_e)\pi^e_{ij}F_{bc}^{ij} = F_{ab}^{ij} \pi^b_{ij}
\end{equation}
Next, modulo $T^a_{ij}=G_{ij}=0$ we have 
$C_a=F_{ab}^{ij} \pi^b_{ij}-A_a^{ij} \hat{G}_{ij}$ generating 
spatial diffeomorphisms on $(A_a^{ij},\pi^a_{ij})$. Since $\hat{P}_a^i, P_i$ 
have already been stabilised, we can add terms linear in $\hat{P}^a_i, P_i$ to $C_a$
so that the resulting constraint $\hat{C}_a$ generates spatial 
diffeomorphisms also on the variables $e^i_a, P^a_i, e^i, P_i$. Since all the constraints are tensor densities,
the constraint $\hat{C}_a$ and thus $C_a$ is already stabilised. \\
\\
Classification of the constraints:\\
Since all constraints are independent of $A^{ij}_t, e^i_t$, both $\pi_{ij}, P_i$ are first class constraints. \\
On the other hand,
since all constraints either are invariant or covariant under Gauss transformations and spatial diffeomorphisms generated by $\hat{G}_{ij}$ and $\hat{C}_a$ respectively, both $\hat{G}_{ij}$ and $\hat{C}_a$ are first class. \\
The constraints $P^a_i, T^a_{ij}$ are second class. To see this we need to show that the matrix
\begin{equation}\label{Matrix}
\Delta^{ab}_{ij}:=\{P^a_i,T^b_j\}=-\epsilon^{abc} \epsilon_{ijk} e_c^k
\end{equation}
is invertible as a symmetric (under $(a,i) \leftrightarrow (b,j)$) $9\times 9$ matrix, where $T^b_j:= T^b_{kl} \epsilon_{jkl}/2$. It is easy to verify that
\begin{equation}
\Delta^{ij}_{ab}=-\frac{3}{4} \epsilon^{ijk}\epsilon_{abc} e^c_k +\frac{1}{2}\det(e)^{-1} e^{(i}_a e^{j)}_b
\end{equation}
is the inverse of (\ref{Matrix}), i.e., $\Delta^{ab}_{ij} \Delta^{jk}_{bc}=\delta^a_c \delta^k_i$ and hence $P^a_i, T^a_{ij}$ form a second class pair.
Therefore, we have $48- (2 \cdot 12 + 18)=6$ degrees of freedom. This means that the action of \cite{0a} cannot be a Lagrangian origin of the 
Hamiltonian U(1)$^3$ theory which has 4 propagating degrees of freedom. Note that just as in \cite{10} the 
Hamiltonian constraint $C=F_{ab}^j\epsilon_{jkl} E^a_k E^b_l$ does not appear as a secondary constraint.

\end{appendix}

\end{document}